\documentclass{iopjournal}
\pdfoutput=1 
\usepackage[numbers]{natbib}           
\usepackage{physics}
\usepackage[symbol]{footmisc}
\usepackage{graphicx}   
\usepackage{subcaption}
\usepackage{makecell}
\usepackage{multirow}
\usepackage{hyperref}
\usepackage{tikz}
\usepackage[table]{xcolor}
\usepackage{colortbl}
\usepackage{braket}
\usepackage{color}
\usepackage{amsmath}
\usepackage{array}
\usepackage{caption}
\begin{document}

\renewcommand{\thefootnote}{\fnsymbol{footnote}}
\title{Gravity induced entanglement of multiple massive particles with large spin}
\author{Kai Li$^{1,2}$, Yi Ling$^{1,2}$, Zhangping Yu$^{1,2,}$\footnote{Corresponding author.}}

\affil{$^1$Institute of High Energy
    Physics, Chinese Academy of Sciences, Beijing 100049, China}
    
\affil{$^2$School of Physics, University of Chinese Academy of Sciences, Beijing 100049, China}

\email{lik@ihep.ac.cn, lingy@ihep.ac.cn, yuzp@ihep.ac.cn}

\renewcommand{\thefootnote}{\arabic{footnote}}
\setcounter{footnote}{0}

\begin{abstract}
We investigate the generation rate of the quantum entanglement in a system composed of multiple massive particles with large spin, where the mass of a single particle can be split into multiple trajectories by a generalized Stern-Gerlach interferometer. Taking the coherent spin states (CSS) as the initial state and considering the gravitational interaction due to Newtonian potential, we compute the generation rate of the entanglement for different configurations of the setup. Explicitly, the optimal polar angles of the spin are found numerically for systems with three and four particles, respectively. We conclude that the amount of the entanglement increases with the number of particles as well as the spin, and the configuration of the prism with a particle at the center generates the best rate of the entanglement. 
\end{abstract}

\newpage
\section{Introduction}
One of the most fundamental problems in modern physics is to diagnose whether the nature of gravity is quantum or classical. Without doubt, the answer would provide significant implications for our understanding of the structure of spacetime and the universe. Unfortunately, the direct detection of quantum effects of gravity remains a formidable challenge due to the extreme weakness of the gravitational interaction relative to other fundamental forces. The naive estimation indicates that the energy level sensitive to the quantum effects of gravity is far beyond the current capability of experiments. Nevertheless, recently a novel strategy based on quantum entanglement has been proposed to experimentally test whether gravity acts as a quantum entity at low energy level \cite{Bose:2017nin, Marletto:2017kzi}. The key idea is to consider two massive particles in the superposition states of position, which can be  generated by two Stern-Gerlach (SG) devices. Suppose initially the system starts in a product state, and during the evolution, these two particles interact with each other solely through gravitational interaction. If quantum entanglement is generated in the final state through this interaction, then gravity, as the exclusive medium between the two particles, must be quantum, as Local Operations and Classical Communication (LOCC) \cite{Plenio:2007zz} cannot generate entanglement from product states. This approach is referred to as ``quantum gravity-induced entanglement of masses (QGEM)". Subsequently, this strategy has been employed to investigate relevant problems, including testing the discreteness of time \cite{Christodoulou:2018xiy}, seeking evidence for quantum superposition of geometries \cite{Christodoulou:2018cmk}, testing gravity-induced reduction of quantum states \cite{Howl:2018qdl}, probing massless and massive gravitons \cite{Elahi:2023ozf}, and validating the weak equivalence principle \cite{Bose:2022czr}. Additional relevant work on this topic can be found in the literature \cite{Belenchia:2018szb, Carney:2018ofe, Marshman:2019sne, Hanif:2023fto, Ghosal:2023vwo, Danielson:2021egj}. Furthermore, inspired by this strategy, alternative strategies have been proposed for testing the nature of gravity at low energy level, such as the exploration by BEC \cite{Howl:2018qdl}, non-Gaussianity \cite{Howl:2020isj}, spacetime diffusion \cite{Oppenheim:2022xjr}, and LOCC but without entanglement \cite{Lami:2023gmz}.

The main challenge in implementing this scheme experimentally is to sustain massive particles in a position superposition for long enough to generate detectable entanglement. Preparing the superposition state for a large massive particle is inherently difficult. Moreover, the gravitational effects from both the environment and the particles themselves can lead to decoherence \cite{Danielson:2022tdw, Danielson:2022sga, Danielson:2024yru, Miki:2020hvg, Moustos:2024ymw}. Given the difficulty of extending the lifetime of superposition states, another feasible approach is to increase the generation rate of entanglement such that the entanglement becomes large enough to be detected within the lifetime of the particles in superposition states. With this success then one may further relax the requirement on the particle mass and thus make the experiment more feasible. To this end, numerous attempts have been made to enhance the generation rate of entanglement. In Ref.\cite{Nguyen:2019huk}, it is found that the entanglement entropy could be increased if the relative position of the particles is rearranged from a linear configuration to a parallel configuration. The setup of multiple particles with different configurations has been explored in Refs.\cite{Schut:2021svd} and \cite{Li:2022yiy}. Furthermore, it is proposed in Ref.\cite{Pedernales:2021dja} to add a classical macroscopic particle as a mediator to enhance gravitational interaction, while in Ref.\cite{vandeKamp:2020rqh} it is proposed to apply a Casimir shield to reduce the spacing between particles. Other improvements on the measurement  can be found in Refs.\cite{Braccini:2024fey, Zhou:2024voj, Lantano:2024usg,Marshman:2023nkh, Schut:2023eux, Schut:2023hsy}.

Increasing the number of particles $N$ is an effective way to enhance the generation rate of entanglement.  It is found in Ref.\cite{Li:2022yiy} that the prism configuration with a particle at the center achieves the fastest rate of entanglement generation. Furthermore, utilizing the configuration with seven particles requires only half of the time to reach the same amount of entanglement as the configuration with three particles. Recently, an alternative way has been proposed to enhance the entanglement by considering particles with large spins \cite{Braccini:2023eyc}. In the original QGEM configuration, spin $1/2$ particles are considered. Since the maximal entanglement is limited by the dimension of the Hilbert space, the amount of entanglement  entropy is bounded by $\ln2$. However, if particles with spin $j$ are employed, the upper bound of entanglement increases to $\ln(2j+1)$. The entanglement of two particles with large spin has been investigated in Ref.\cite{Braccini:2023eyc}, indicating that large spins can significantly enhance both the entanglement generation rate and the maximal value of entanglement. Inspired by the above work, we intend to investigate the rate of entanglement generation in the case of multiple particles with large spins. We will consider the system consisting of multiple particles with large spin, and figure out all the possible configurations for the setup,  and then compute the generation rate of the entanglement for different configurations. Explicitly, the optimal polar angles of the spin are found numerically for the systems with three and four particles, respectively. By the above analysis we find that the amount of the entanglement increases with the number of particles as well as the spin, and the configuration of the prism with a particle at the center generates the best rate of the entanglement.

The paper is organized as follows. The general setup for multiple massive particles with large spin is presented in the next section, with details on the evolution of the system driven by the gravitational potential. The generation rate of entanglement for the system with three particles is investigated in Section \ref{sec3}. We numerically compute the entanglement entropy for the system up to $j=5$ and figure out the optimal polar angles of the spin. It turns out that the rate of entanglement generation is significantly improved in comparison with the system with two particles. In Section \ref{sec4} we numerically compute the entanglement entropy for the system with four particles up to $j=2$, and the rules for obtaining the optimal polar angles of the spin are obtained. In Section \ref{sec:decoherence}, we investigate the impact of decoherence on the experiment. Our conclusions and discussions are given in the last Section.

\section{The general setup for multiple massive particles with large spin}

In this section, we present the general setup for multiple massive particles with large spin. We begin by introducing the superposition state for a particle with spin $j$, which may be intuitively described by $2j+1$ semiclassical trajectories generated by the Generalized Stern-Gerlach (CSG) interferometer, as proposed in \cite{Braccini:2023eyc}. Then we consider a system consisting of multiple particles, each with $2j+1$ semiclassical trajectories, interacting with one another through Newtonian potential. We outline the logic line for the computation of entanglement entropy between one specified particle with the other particles in this system, which serves as the basis for witnessing gravity-induced entanglement in QGEM experiment. 

The generalized Stern-Gerlach interferometer, which splits the mass with spin $j$ into $2j + 1$ trajectories, was firstly explored in Ref.\cite{Braccini:2023eyc}. The protocol of the process can be described as follows:

(1) \textit{Initial State Preparation:} The initial state is prepared as a tensor product of spin state and position state: $\ket{\psi(t=0)} = \ket{\psi_S}\otimes \ket{\psi_x} = \left(\sum\limits_{m=-j}^j c_m \ket{m}\right)\otimes\ket{0}$, where $\ket{\psi_S}$ is a specific spin state, and $c_m$ are the coefficients in the Dicke basis $\left\{\ket{j,m}\right\}$ with $m \in [-j, j]$, and $\ket{\psi_x} = \ket{0}$ represents the position ground state. The specific spin state $\ket{\psi_S}$ can be created by applying secondary magnetic fields to the spin ground state $\ket{m=-j}$. Three families of spin states are discussed in Ref.\cite{Braccini:2023eyc}: Coherent Spin States (CSS), a superposition of CSS, and Squeezed Spin States (SSS). It was found that there was no significant difference between using the last two families of spin states and using the first one; thus, for simplicity, we utilize Coherent Spin States (CSS) in this paper. A CSS is defined as the state resulting from an arbitrary rotation of the spin ground state $\ket{-j}$, and for the CSS state with a specific direction,
\begin{align}
    \ket{\psi_{CSS}}=\ket{\phi,\theta} & :=\mathcal{N} e^{\mu J_-} \ket{m=-j}\nonumber\label{eq:CSS}             \\
                                       & =\mathcal{N} \sum_{m=-j}^j  \mu^{j+m}\sqrt{\frac{2j!}{(j+m)!(j-m)!}}\ket{m},
\end{align}
where $\mathcal{N}=\left(1-\abs{\mu}^2\right)^{-j}$ is the normalization factor with $\mu=e^{i\phi} \tan{\theta/2}$, and $\phi$, $\theta$ are understood as the phase in the $xy$-plane  and the azimuth angle with respect to the $z$-axis, respectively.

(2) \textit{Splitting Process:} The splitting process is described by a Hamiltonian
\begin{align}
    H=\hbar \omega_{M} a^{\dagger} a-\hbar g J_{z}\left(a+a^{\dagger}\right) ,
\end{align}
where $\omega_M$ is the frequency of the quantum harmonic oscillator in which the mass is trapped, and the coupling constant $g$ between spin and position for the mass $M$ is given by
\begin{align}
    g=\tilde{g} \mu_{B} \sqrt{\frac{1}{2 \hbar M \omega_{M}}}\left(\partial_{x} B\right),
\end{align}
where $\mu_{B}$ is the Bohr magneton and $\tilde{g}$ is the Lande g-factor. The derivation of the time evolution can be found in Appendix B of Ref.\cite{Braccini:2023eyc}, and the the quantum state at time $t$ is given by
\begin{align}
    |\psi(t)\rangle=\sum_{m=-j}^{j} c_{m} e^{i \frac{g^{2}}{\omega_{M}^{2}} m^{2}\left(\omega_{M} t-\sin \left(\omega_{M} t\right)\right)}|m\rangle \otimes\left|\alpha_{m}(t)\right\rangle,
\end{align}
where the position coherent state is given by
\begin{align}
    \alpha_{m}(t)=m \frac{g}{\omega_{M}}\left(1-e^{-i \omega_{M} t}\right) .
\end{align}
The maximal displacement between adjacent trajectories is achieved when $t_s = \pi / \omega_M$, and the splitting
\begin{align}
    \Delta x := \expval{x_{m+1}(t_s)} - \expval{x_m(t_s)} = 2\sqrt{\frac{2\hbar}{M \omega_M}} \frac{g}{\omega_M}
\end{align}
is independent of $m$, where $\expval{x_{m}(t_s)} = \expval{X}{\alpha_m(t_s)}$ is the position of the $m$-th trajectory. At the time $t_s$, the state is given by
\begin{align}
    \left|\psi\left(t_{s}\right)\right\rangle=\sum_{m=-j}^{j} c_{m} e^{i \pi \frac{g^{2}}{\omega_{M}^{2}} m^{2}}|m\rangle \otimes\left|\alpha_{m}=m \Delta x\right\rangle.
\end{align}

(3) \textit{Recombination and Measurements:} After the interaction, $2j+1$ trajectories can be recombined through the inverse process of splitting, and the final state is given by
\begin{align}
    \left|\psi\left(2 t_{s}\right)\right\rangle=\left(\sum_{m=-j}^{j} c_{m} e^{i 2 \pi \frac{g^{2}}{\omega_{M}^{2}} m^{2}}|m\rangle\right) \otimes|0\rangle.
\end{align}
The spatial degrees of freedom and spin degrees of freedom of the final state are separated after recombination, allowing for general spin measurements to be performed on the spin component embedded in the mass.

Next we consider a system composed of $N$ massive particles with identical spin $j$ interacting via gravity due to
Newtonian potential, each of which splits into $2j + 1$ trajectories. The initial state of the system is given by $\ket{\Psi(t_s)} =\prod\limits_{i=1}^{N}\otimes\ket{\psi_i(t_s)}$, where
\begin{align}
    \ket{\psi_i(t_s)} = \sum_{m=-j}^{j} c_{m}(\phi_i, \theta_i) e^{i \pi \frac{g^{2}}{\omega_{M}^{2}} m^{2}} \ket{m} \otimes\ket{x_i(m)}.
\end{align}
$x_i(m)$ is the position of the $i$-th particle along the $m$-th trajectory, and its specific value depends on the configuration of arranging $N$ particles. In this paper, we will consider several different configurations. Note that each particle is created by splitting an initial state $\ket{\psi_i(t=0)} = \ket{\psi_{CSS}}\otimes \ket{(x_0)_i}$, but the orientation of the CSS can be different. However, it can be seen from Eq.(\ref{eq:CSS}) that $\phi_i$ only contributes an overall phase factor, which does not affect the calculation of entanglement entropy. Therefore, without loss of generality, we set $\phi_i=0$, for $i=1,\ldots,N$.

The evolution of the system is described by the Hamiltonian \cite{Schut:2021svd,Li:2022yiy}
\begin{equation} \label{eq:Hamiltonian}
    \hat{H} = \sum_{1 \le k < l \le N} \hat{V}_{kl},
\end{equation}
where $\hat{V}_{kl}$ is the gravitational potential between the $k$-th particle and the $l$-th particle, and
\begin{align}
    \left(\hat{V}_{kl}\right)_{m,n} = - \frac{G M^2}{R\left(x_k(m), x_l(n)\right)},
\end{align}
where $R(x_k(m), x_l(n))$ is the distance between the $k$-th particle along the $m$-th trajectory and the $l$-th particle along the $n$-th trajectory, with $m,n \in [-j,j]$. For various configurations under consideration, we present the specific expressions for  $R(x_k(m), x_l(n))$ in Appendix \ref{appendix:A}. The Hamiltonian in Eq.~(\ref{eq:Hamiltonian}) consists solely of the interaction term, as the kinetic energy contribution $p^2/2m$ is negligible for the small particle momenta in the experiment. Consequently, no entanglement is generated in the absence of gravity.

It is straightforward to obtain the state of the system after $t$ seconds of interaction:
\begin{align}
    \ket{\Psi(t_s + t)} & = e^{-i \hat{H}t} \ket{\Psi(t_s)}\nonumber                                                                                                                                                 \\
                        & =\sum_{m_1,\ldots,m_N=-j}^{j} \left[\left(\prod_{i=1}^{N}c_{m_i}(\theta_i)\right)e^{i \pi \frac{g^2}{\omega_M^2}\left(\sum_{i=1}^N m_i^2\right)-i\phi_{m_1 \ldots m_N} t}\right. \nonumber \\
                        & \quad\quad\quad\left.\prod_{i=1}^{N}\left(\ket{m_i}\otimes\ket{x_i(m_i)}\right)\right],
\end{align}
where the phase $\phi$ is determined by the Newtonian potential as
\begin{align}
    \phi_{m_1 \ldots m_N} = - \sum_{1 \le k < l \le N} \frac{G m^2}{R(x_k(m_i), x_l(m_j))}.
\end{align}
After the recombination process, the final state is given by
\begin{align}
    \label{eq:finalState}
    \ket{\Psi(2t_s + t)} & =\left[\sum_{m_1,\ldots,m_N=-j}^{j} \left(\prod_{i=1}^{N}c_{m_i}(\theta_i)\right)e^{i \pi \frac{2g^2}{\omega_M^2}\left(\sum_{i=1}^N m_i^2\right)-i\phi_{m_1 \ldots m_N} t}\prod_{i=1}^{N}\ket{m_i}\right]\nonumber \\
                         & \quad\otimes\prod_{i=1}^{N}\ket{(x_0)_i}.
\end{align}

Using Eq.~(\ref{eq:finalState}), we may compute the entanglement among the particles. Since the spatial degrees of freedom and spin degrees of freedom are separated, we only need to consider the entanglement between the spins. We intend to emphasize that although the final state appears to involve the entanglement of spins only, this entanglement is, in fact, generated entirely by the gravitational interaction, as the Hamiltonian does not include any spin-spin interaction term among the particles.

Since the entire system is in a pure state, the entanglement entropy between two subsystems is quantified by the von Neumann entropy \cite{von1955mathematical}: the entanglement between particle $i$ and the other particles can be measured by
\begin{equation}
    S_i = S(\rho_i) = - \Tr\qty(\rho_i \ln \rho_i) = -\sum_j \lambda_j \ln \lambda_j,
\end{equation}
where $\rho_i = \Tr_{1, \cdots, \hat{i}, \cdots, N}(\op{\Psi(2t_s + t)})$ is the reduced density matrix of the particle $i$, and $\lambda_j$ are the eigenvalues of $\rho_i$.

However, the von Neumann entropy, is not a valid measure of entanglement when the global system is in a mixed state. This occurs, for instance, when describing the test masses as an open quantum system (i.e., including decoherence). The von Neumann entropy does not distinguish between quantum correlations arising from entanglement and classical correlations due to the statistical mixture of states. A non-zero von Neumann entropy can exist simply because the subsystem is part of a classically correlated state (a mixed state), even in the complete absence of entanglement. 

For a system in a mixed state, a widely used measure of entanglement is negativity \cite{PhysRevA.58.883,PhysRevLett.95.090503}. It is defined as the absolute value of the sum of the negative eigenvalues of the partial transposed density matrix. Specifically, once the density matrix $\rho$ of the global system is obtained, the negativity between particle $i$ and the remaining particles is given by $\mathcal{W}_i = \abs{\sum_{\lambda < 0 } \lambda}$, where $\lambda$ are the eigenvalues of the partial transposed density matrix $\rho^{PT_i}$. A non-zero negativity certifies the presence of quantum entanglement (although zero negativity does not necessarily imply the absence of entanglement). Therefore, when studying the effects of decoherence, we consider the negativity as an alternative measure of the entanglement, and the details are presented in Section \ref{sec:decoherence}. 

\begin{figure}[h]
    \centering
    \begin{subfigure}[b]{0.9\textwidth}
        \centering
        \includegraphics[width=0.8\textwidth]{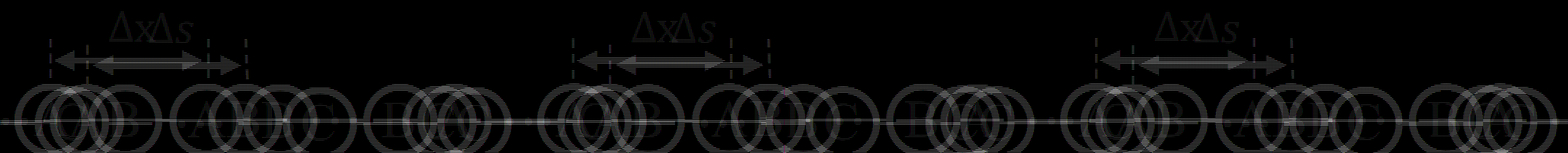} 
        \caption{Linear}
        \label{fig:single_image}
    \end{subfigure}

    \vfill

    \begin{subfigure}[b]{0.25\textwidth}
        \centering
        \includegraphics[width=\textwidth]{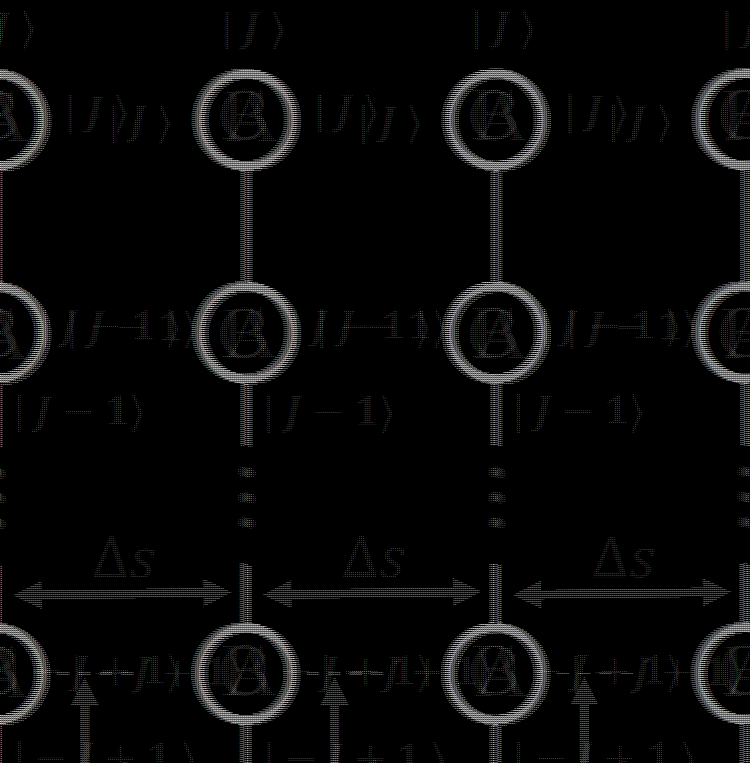} 
        \caption{Parallel}
        \label{fig:second_row_left}
    \end{subfigure}
    \hspace{0.1\textwidth} 
    \begin{subfigure}[b]{0.25\textwidth}
        \centering
        \includegraphics[width=\textwidth]{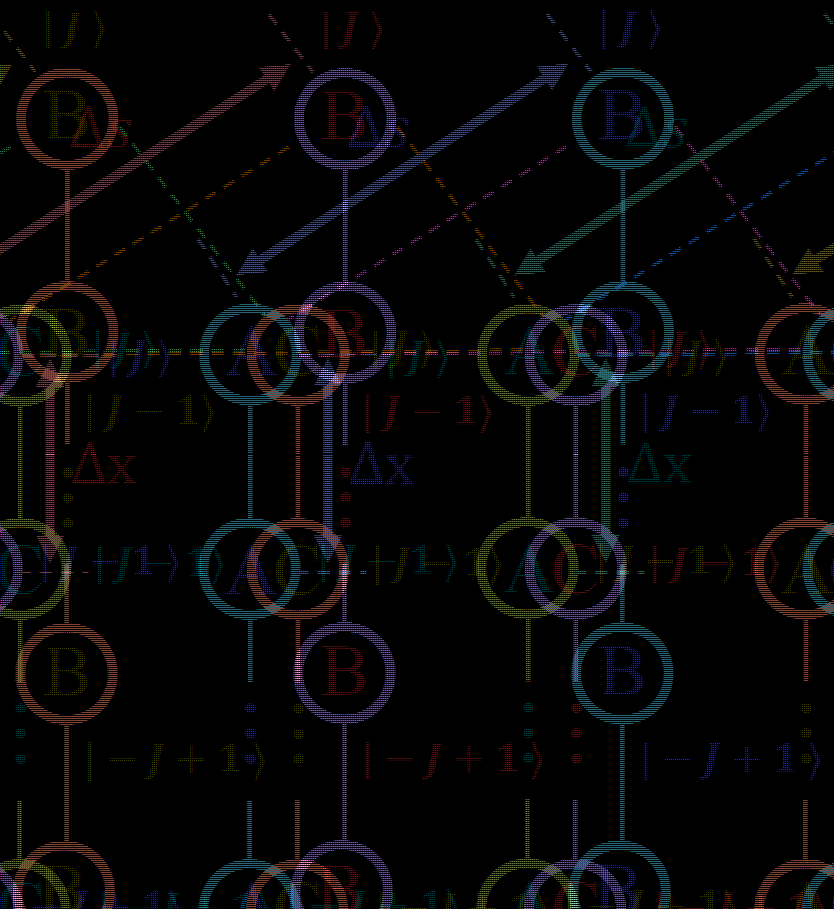} 
        \caption{Prism}
        \label{fig:second_row_right}
    \end{subfigure}

    \vfill

    \begin{subfigure}[b]{0.35\textwidth}
        \centering
        \includegraphics[width=\textwidth]{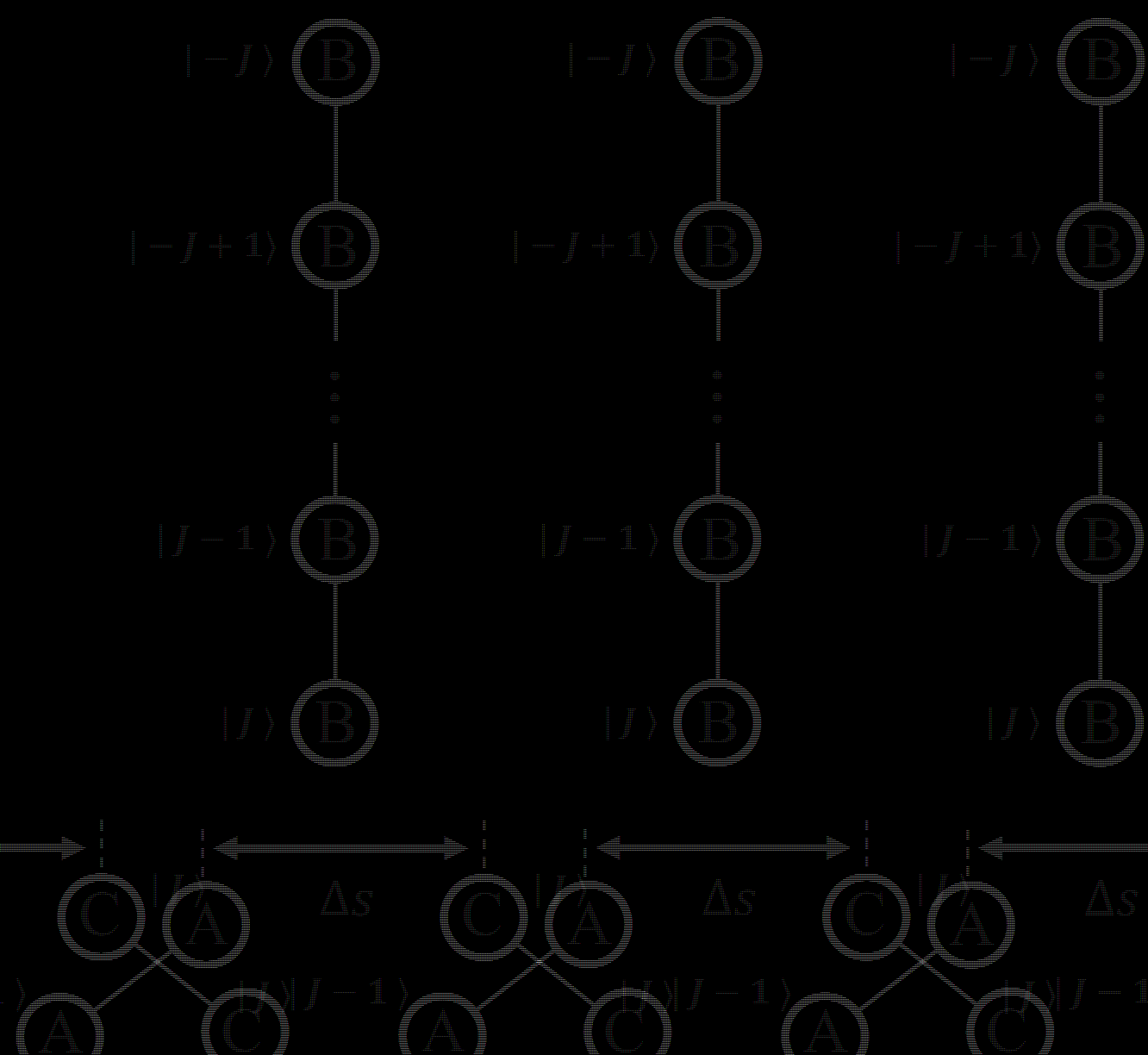} 
        \caption{Star}
        \label{fig:third_row_left}
    \end{subfigure}
    \hspace{0.04\textwidth} 
    \begin{subfigure}[b]{0.32\textwidth}
        \centering
        \includegraphics[width=\textwidth]{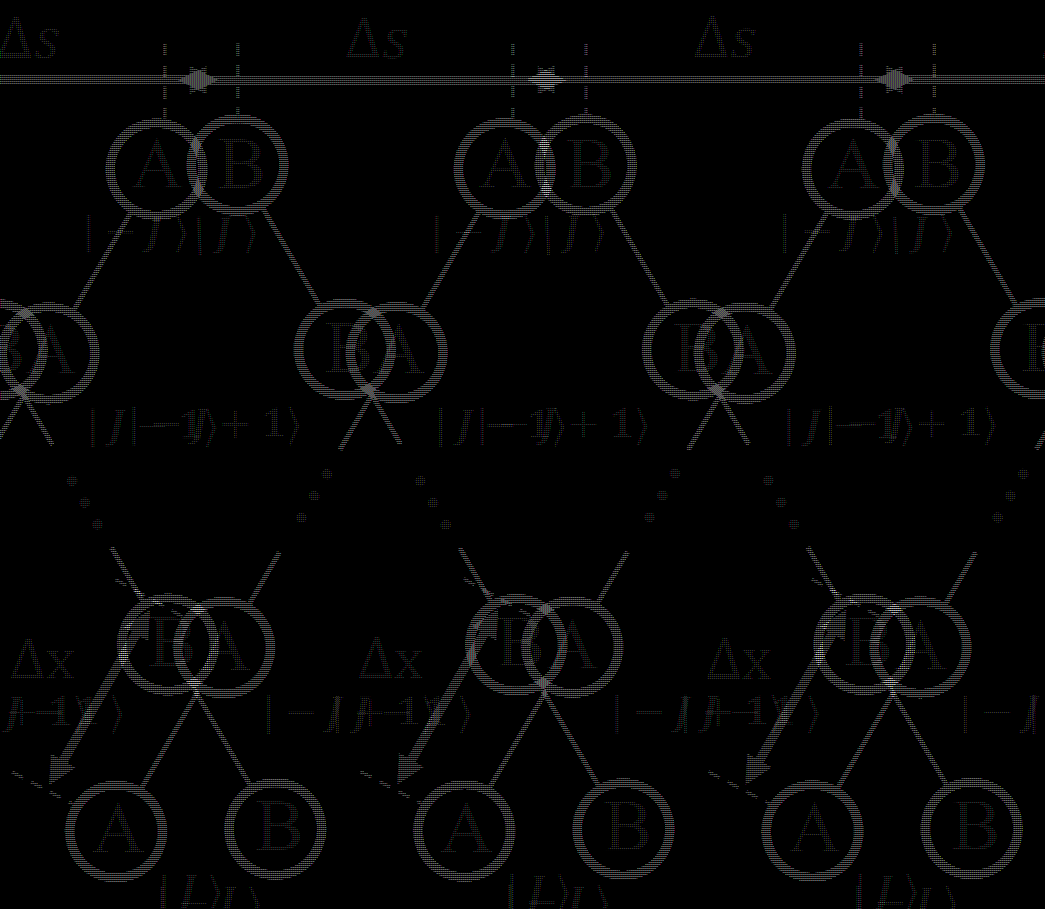} 
        \caption{Polygon}
        \label{fig:third_row_right}
    \end{subfigure}
    \captionsetup{justification=raggedright, singlelinecheck=false}
    \caption{The allowable configurations for the system consisting of three particles with large spins.}
    \label{fig:n3setup}
\end{figure}

\section{The gravity induced entanglement in the system with three  particles with large spin}\label{sec3}
In this section we investigate the generation rate of entanglement  for the system composed of  three massive particles with spin $j$.  Due to the symmetry, the distinct configurations of arranging these three large-spin particles are illustrated in Figure (\ref{fig:n3setup}), which include the \textit{Linear}, \textit{Polygon}, \textit{Star}, \textit{Prism}, and \textit{Parallel} configurations\cite{Li:2022yiy}. In the case of $j=1/2$, it is found in \cite{Li:2022yiy} that  both the \textit{Parallel} and \textit{Prism} configurations  exhibit the fastest entanglement generation rates compared to other configurations shown in Fig. (\ref{fig:n3setup}). Here we extend the study for large spin.  

For an $N$-particle system, there are $N$ parameters $\theta_i$, where $i = 1, \dots, N$. We need to optimize the polar angles $\theta_i$, as these have a more significant effect on the entanglement generation. Given the complexity of the problem, an analytical approach is challenging; thus, we employ a numerical grid search method to figure out  the optimal values of $\theta_i$ that maximize the entanglement entropy.

To compare with the results in previous work on two-particle system of large spins \cite{Braccini:2023eyc}, we adopt the experimental configuration proposed in \cite{Bose:2017nin}, with the following specification for parameters: interaction time \(\tau = 2 \, \text{s}\), particle mass \(M_i \approx 10^{-14} \, \text{kg}\), and spatial separation \(\Delta x \approx 250 \, \mu\text{m}\). Additionally, the effects of Casimir screening \cite{vandeKamp:2020rqh} are incorporated, which reduces the separation between particles to $\Delta s = 50 \, \mu\text{m}$.

\begin{table}[!ht]
\centering
\caption{The Von Neumann entanglement entropy of different configurations with $N = 3$ and $t = 2$s}
\begin{tabular}{|c|c|c|}
\hline
Configuration & \makecell{Entanglement entropy\\($J=2$)} & \makecell{Entanglement entropy\\($J=5$)} \\
\hline
Linear   & 0.540   & 0.538   \\
\hline
Polygon        &    0.566    &     0.563    \\
\hline
Star   & 0.653   & 0.655  \\
\hline
\textbf{Parallel or Prism}        &     \textbf{1.367}    &      \textbf{1.647}   \\
\hline
\end{tabular}
\label{tab:tablen3config}
\end{table}

For clarity, we present the results for \(j = 2\) and \(j = 5\) here, while the detailed data for other values of \(j\) and the corresponding optimal values of \(\theta_i\) are provided in Appendix \ref{appendix:B}. The von Neumann entropy, \(S_2\), is evaluated at the time \(t = 2\) s. From Table (\ref{tab:tablen3config}), we observe that  the maximal entropy increases with the spin indeed, thus the larger spin is beneficial to the entanglement generation. Moreover, given the spin $j$,  the maximal entropy for the system with three particles is greater than that for the system with two particles at the same evolution time \cite{Braccini:2023eyc}, thus adding more particles is also beneficial to the entanglement generation.

We summarize the concrete rules for optimal angles of maximum entanglement entropy for each configuration in Appendix \ref{appendix:B}. We notice that the pattern of the optimal \(\theta_i\) for the Linear and Polygon configurations is quite similar. In both cases, the maximal entanglement entropy appears when \(\theta_A + \theta_B = \pi\) or \(\theta_B + \theta_C = \pi\), and the remaining angle, \(\theta_C\) or \(\theta_A\), has a relatively small effect on the result. For the Star configuration, the maximum of entropy is achieved when \(\theta_A = \theta_B = \theta_C\). As the value of \(j\) increases, the optimal \(\theta_i\) gradually decreases. Furthermore, for the three-particle case, we find that the optimal \(\theta_i\) and the maximal entropy are identical for both the Parallel and Prism configurations. Especially, we list the rules of these two configurations in Table \ref{tab:tablen3parallelrules}. The optimal \(\theta_i\) for these configurations satisfies \(\theta_A + \theta_C = \pi\) and \(\theta_B = \pi / 2\). Notably, the Prism/Parallel configuration achieves the largest entanglement entropy among all the configurations considered. This conclusion is the same as that found in the case of $j=1/2$, but the amount of entanglement entropy has been greatly enhanced with the increase of spin $j$.

\begin{table}[!ht]
\centering
\caption{The rules for optimal angles for the maximal entropy of Parallel/Prism configurations with $N=3$}
\begin{tabular}{|c|c|}
\hline
Configuration & Rules for optimal angles  \\
\hline
$J=1/2$ or $1$  & $\theta_A=\theta_B=\theta_C=\pi/2$      \\
\hline
$J>1$        &    \makecell{$\theta_B=\pi/2$\\ $\theta_A +\theta_C=\pi$}        \\ 
\hline
\end{tabular}
\label{tab:tablen3parallelrules}
\end{table}

Next, to explicitly demonstrate the change of the entanglement entropy with the angles $\theta_i$, we perform a contour plot for the entanglement entropy on $(\theta_A, \theta_C)$ plane for different configurations in Fig. (\ref{fig:j2different}), where $\theta_B$ is fixed to be the value presented in Appendix \ref{appendix:B}. From this figure, it is evident that given the spin $j$, the Prism/Parallel configuration yields the largest entanglement entropy. On the other hand, for a given configuration, one finds that the region with relatively large entropy becomes narrow with the increase of spin $j$, while simultaneously the maximum of entropy becomes larger indeed. This indicates that larger spins enhance the entanglement capability of the system, thereby improving its ability to generate and sustain quantum entanglement.

\begin{figure}[htbp]
    \centering
    \begin{subfigure}{0.48\textwidth}
        \centering
        \includegraphics[width=\textwidth]{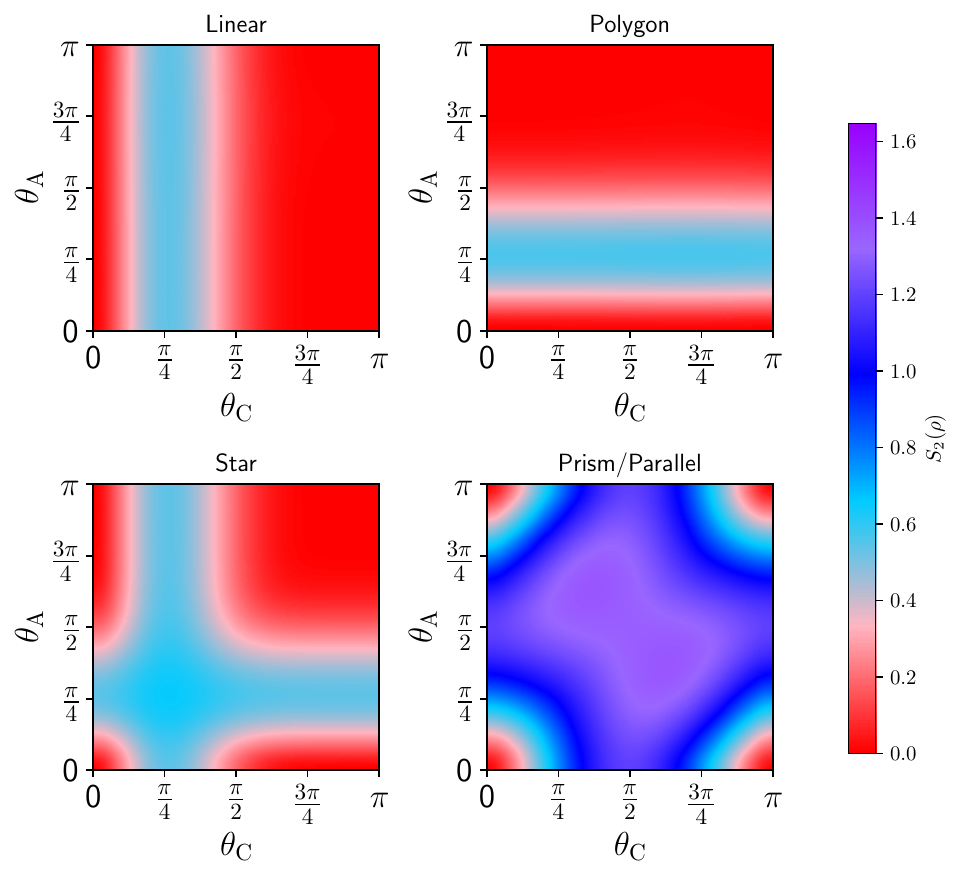}
        \caption{$J=2$}
        \label{fig:j2different1}
    \end{subfigure}
    \hfill 
    \begin{subfigure}{0.48\textwidth}
        \centering
        \includegraphics[width=\textwidth]{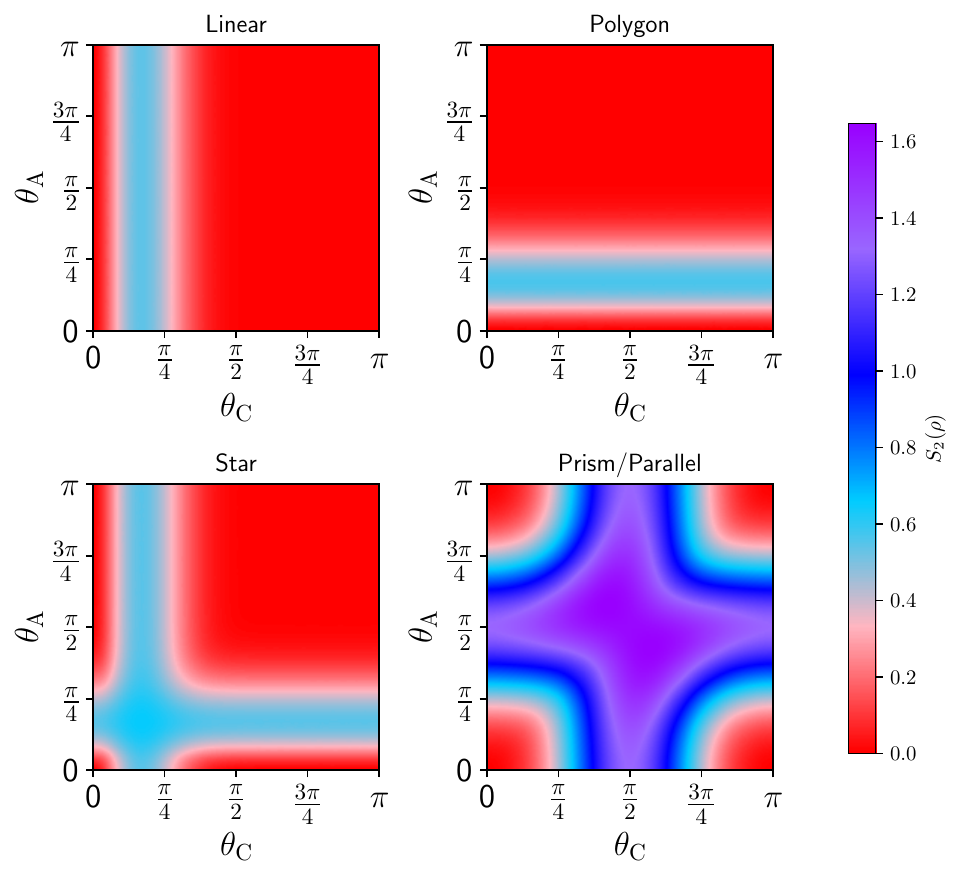}
        \caption{$J=5$}
        \label{fig:j2different2}
    \end{subfigure}
    \captionsetup{justification=raggedright, singlelinecheck=false}
    \caption{The contour plot for the entanglement entropy over $(\theta_A, \theta_C)$ plane for different configurations with three particles at $t=2$s.
    The specific configuration is labeled on the top of each subfigure.} 
    \label{fig:j2different}
\end{figure}

\begin{figure}[htbp]
    \centering
    \begin{subfigure}{0.48\textwidth}
        \centering
        \includegraphics[width=\textwidth]{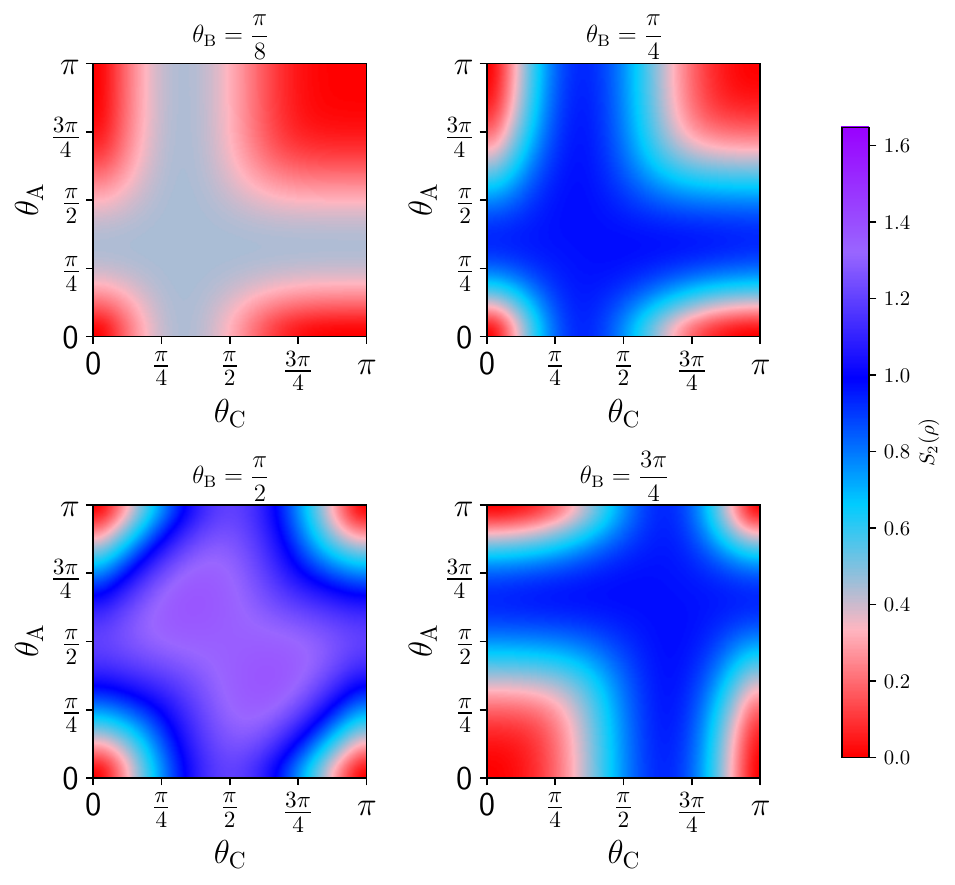}
        \caption{$J=2$}
        \label{fig:n3same1}
    \end{subfigure}
    \hfill 
    \begin{subfigure}{0.48\textwidth}
        \centering
        \includegraphics[width=\textwidth]{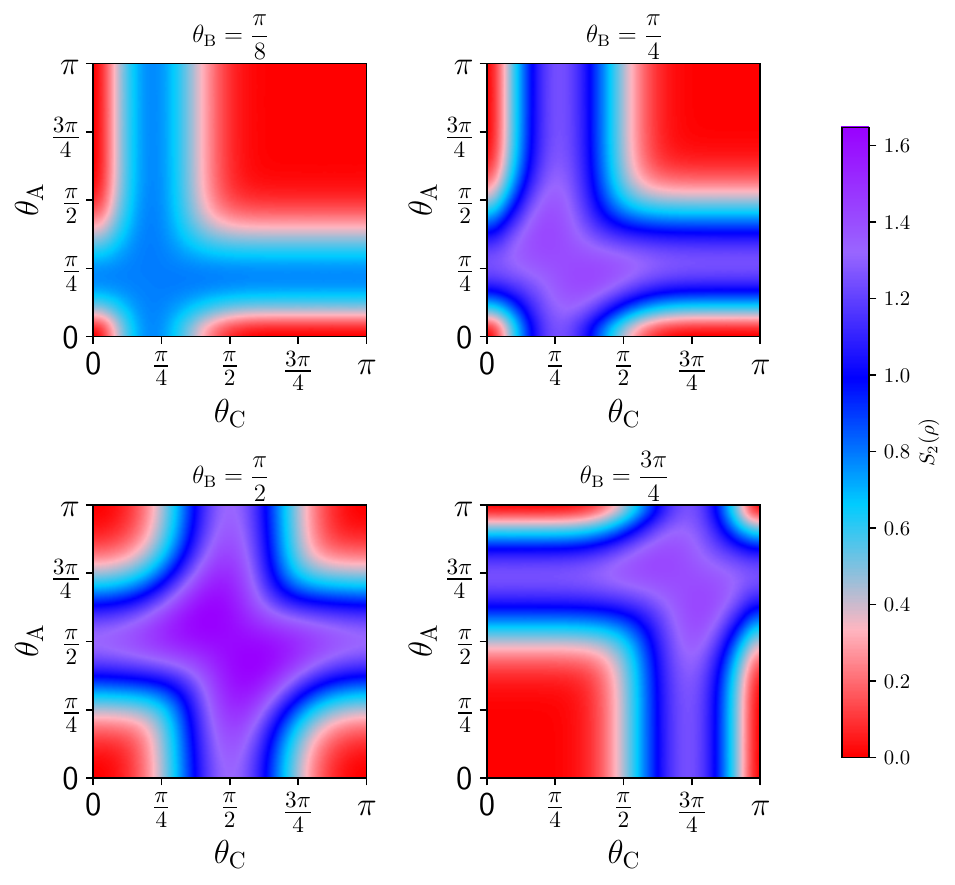}
        \caption{$J=5$}
        \label{fig:n3same2}
    \end{subfigure}
    \captionsetup{justification=raggedright, singlelinecheck=false}
    \caption{The contour plot for the entanglement entropy over $(\theta_A, \theta_C)$ plane for different values of $\theta_B$ in the $Prism/Parallel$ configuration with three particles at $t=2$s.} 
    \label{fig:n3same}
\end{figure}

Next we examine the effect of \(\theta_B\) on the entanglement entropy for the Prism/Parallel configuration, as illustrated in Fig. (\ref{fig:n3same}). It is evident that given the spin $j$, when \(\theta_B\) deviates from \(\pi / 2\), the entropy decreases rapidly. Specifically, when \(\theta_B = 0\) or \(\theta_B = \pi\), particle B is no longer entangled with the other two particles. On the other hand, as the spin \(j\) increases, the reduction of entropy becomes less pronounced. In addition, we notice that there exists a symmetry between  the cases of \(\theta_B\) and \(\pi - \theta_B\), demonstrating the inherent symmetry of the system with respect to \(\theta_B\).

Now we turn to discuss the time evolution of the entanglement entropy for different configurations and different spins. The results are presented in Fig. (\ref{fig:n3time}), while the complete period is illustrated in Appendix \ref{appendix:B}. Fig. (\ref{fig:subn3time1}) presents the evolution of the entanglement entropy for five distinct configurations with spin $j=2$ (dashed lines)  and $ j=5$ (solid lines), while Fig. (\ref{fig:subn3time2}) compares the  evolution of entropy for the system with two particles and the system with three particles. As observed in Fig. (\ref{fig:subn3time1}), for a given spin, the \( Prism/Parallel \) configuration consistently exhibits the highest entanglement entropy among all the configurations analyzed. Fig. (\ref{fig:subn3time2}) further highlights that the entanglement entropy of the system with two particles (represented by the dashed baby blue curve) is notably lower than that of the system with three particles with the same spin (depicted by the solid blue curve). This indicates that increasing the number of particles significantly enhances the entanglement entropy.

\begin{figure}[htbp]
    \centering
    \begin{subfigure}[t]{0.48\textwidth}
        \centering
        \includegraphics[width=\textwidth]{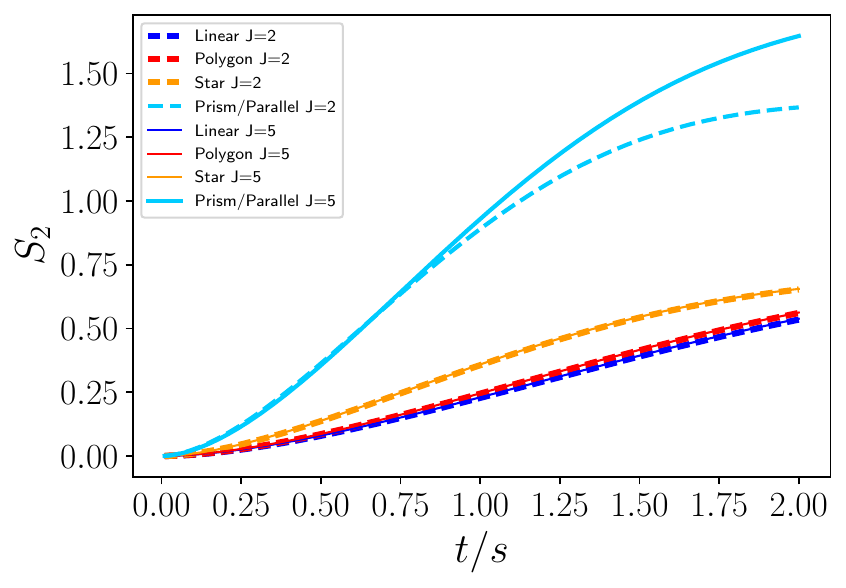}
        \caption{The evolution of different configurations}
        \label{fig:subn3time1}
    \end{subfigure}
    \hfill 
    \begin{subfigure}[t]{0.48\textwidth}
        \centering
        \includegraphics[width=\textwidth]{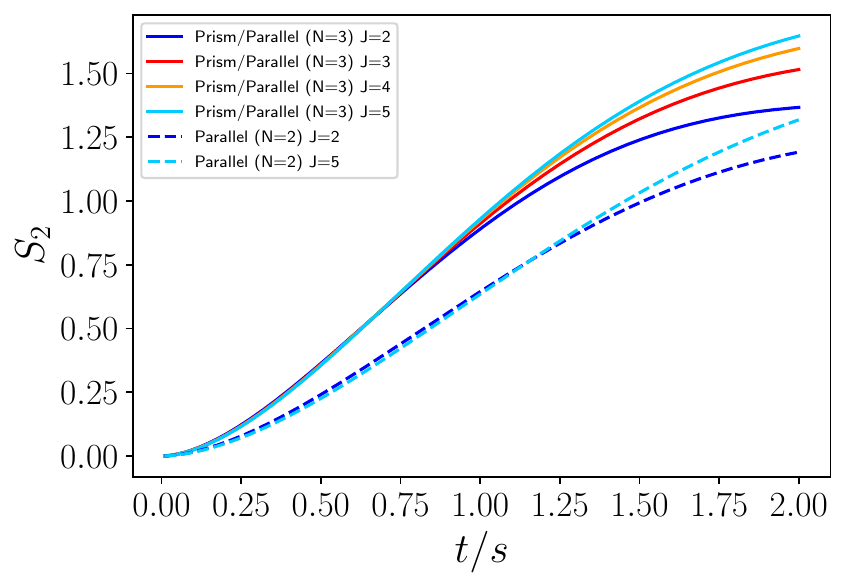}
        \caption{The evolution of different spins and particles}
        \label{fig:subn3time2}
    \end{subfigure}
    \caption{
    The evolution of the entanglement entropy for various configurations and spins.} 
    \label{fig:n3time}
\end{figure}

\section{The gravity induced entanglement in the system with four  particles with large spin}\label{sec4}

\begin{figure}[htbp]
    \centering
    \begin{subfigure}{0.3\textwidth}
        \centering
        \includegraphics[width=\textwidth]{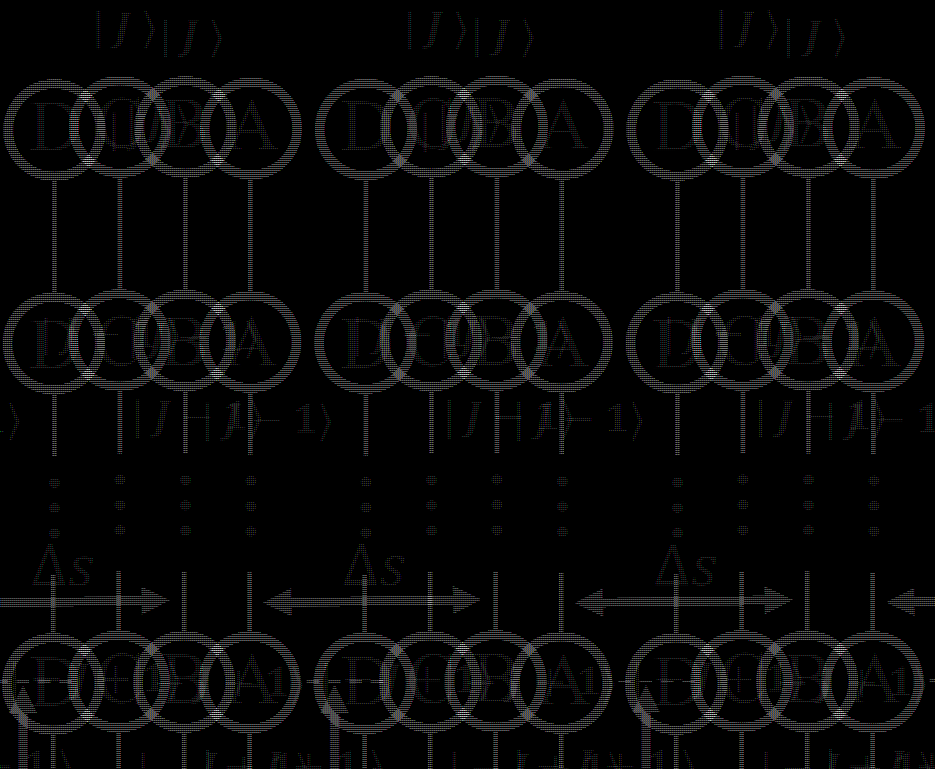}
        \caption{Parallel}
        \label{fig:n4setup1}
    \end{subfigure}
    \hfill 
    \begin{subfigure}{0.3\textwidth}
        \centering
        \includegraphics[width=\textwidth]{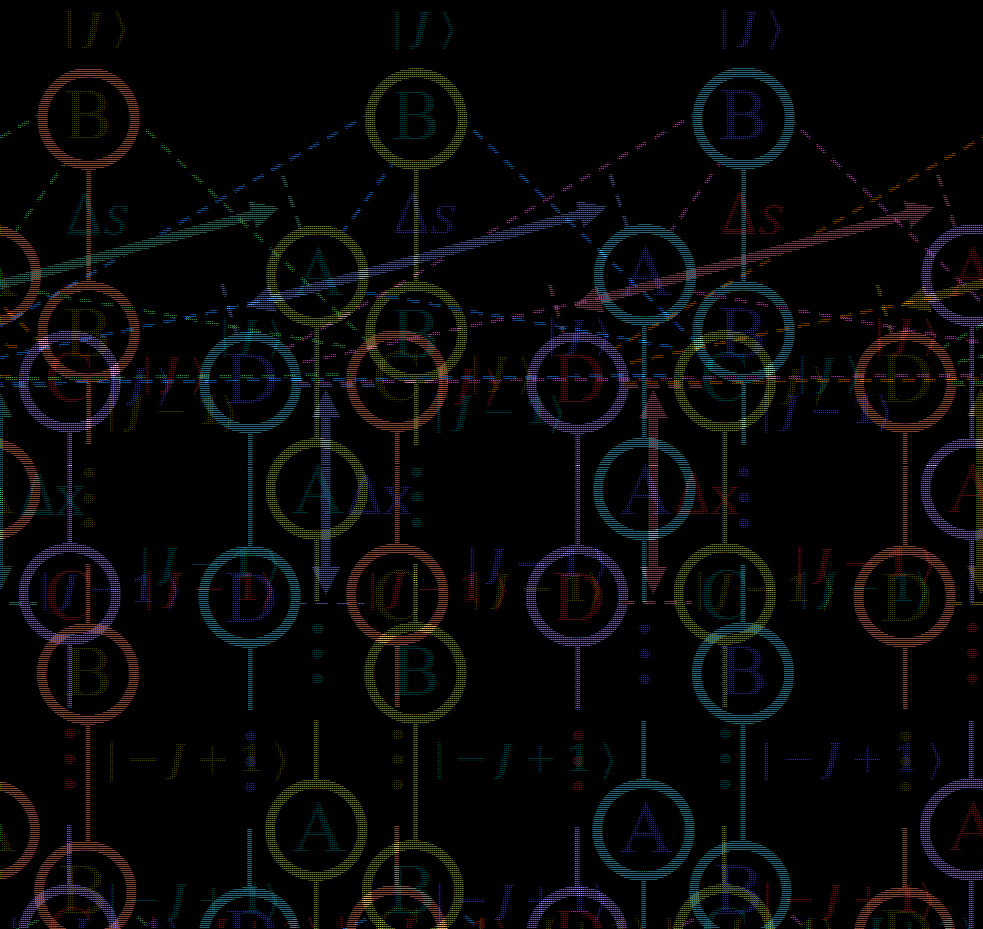}
        \caption{Prism with center}
        \label{fig:n4setup2}
    \end{subfigure}
    \hfill 
    \begin{subfigure}{0.3\textwidth}
        \centering
        \includegraphics[width=\textwidth]{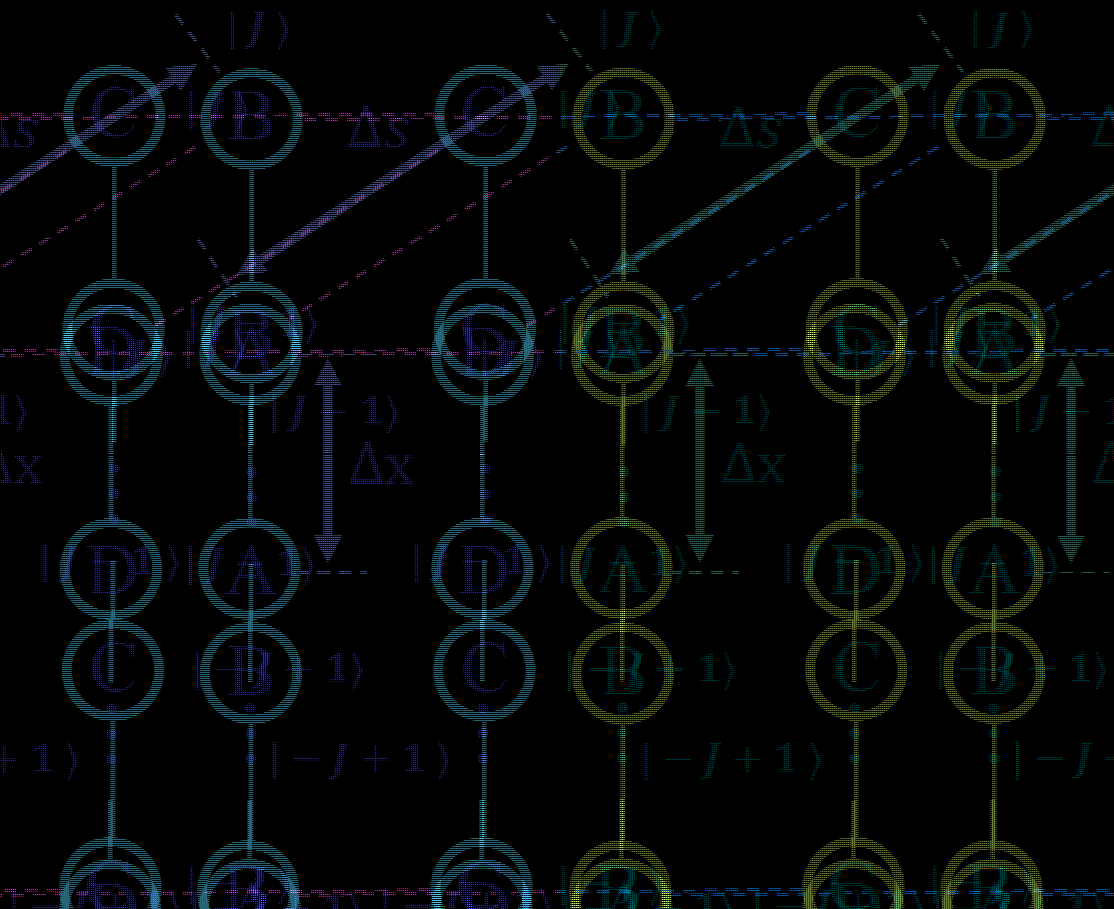}
        \caption{Prism}
        \label{fig:n4setup3}
    \end{subfigure}
    \captionsetup{justification=raggedright, singlelinecheck=false}
    \caption{Three typical configurations for the system consisting of four particles with large spins.
}
    \label{fig:n4setup}
\end{figure} 

In this section, we focus on the system with four particles with large spin $(N=4)$. Unlike the previous section, we restrict our analysis to the \( Parallel \) and \( Prism \) configurations for four particles, as these configurations exhibit the highest efficiency of entanglement generation. For the \( Prism \) configuration with \( N=4 \), there are two possible configurations, as discussed in \cite{Li:2022yiy}. Now extending to particles with large spin, we illustrate the configurations in Fig. (\ref{fig:n4setup}), corresponding to \( Parallel \), \( Prism\ with\ center \), and \( Prism \), respectively.

\begin{table}[!ht]
\centering
\caption{The Von Neumann entanglement entropy of different configurations with $N = 4$ and $t = 2$s}
\begin{tabular}{|c|c|c|}
\hline
Configuration & \makecell{Entanglement entropy\\($J=1$)} & \makecell{Entanglement entropy\\($J=2$)} \\
\hline
Parallel  &  1.036   & 1.377   \\
\hline
Prism        &    1.038    &     1.387    \\
\hline
\textbf{Prism with center}   & \textbf{1.039}   & \textbf{1.399}  \\
\hline
\end{tabular}
\label{tab:tablen4config}
\end{table}

Similar to the case of three particles, we perform a numerical scan over four parameters \( \theta_i \) to figure out the optimal combination that maximizes the entanglement entropy. Since  the computational resources will increase exponentially with the number of particles, in this section we only analyze the cases with spin up to $j=2$. The detailed results and corresponding rules are summarized in Appendix  \ref{appendix:B}. From Table \ref{tab:tablen4config}, we conclude that among these three configurations, the \textit{Prism with center} configuration achieves the maximum entanglement generation rate, while actually they are quite close to each other.

Moreover, from Table (\ref{tab:tablen4rules}), one notices that for smaller spin values, such as $j=1/2$ and $j=1$, the optimal value for all \( \theta_i \) is \( \pi / 2 \). As a matter of fact, this rule also holds for all configurations with three particles, except for the \( Star \) configuration, as also shown in Appendix \ref{appendix:B}. 

For \( j=2 \), the \( Parallel \) and \( Prism \) configurations follow the same rules, namely \( \theta_A + \theta_C = \pi \) and \( \theta_B = \theta_D = \pi / 2\). Additionally, we show that \( \theta_B \), \( \theta_C \), and \( \theta_D \) in the \( Prism\ with\ center \) configuration exhibit cyclic symmetry. 

\begin{table}[!ht]
\centering
\caption{The rules for optimal angles for the maximal entropy of Parallel/Prism/Prism with center configurations with $N=4$}
\begin{tabular}{|c|c|c|}
\hline
Spin & Configuration & Rules for optimal angles \\
\hline
$J=1/2$ or $1$ & \makecell{Parallel, Prism\\or Prism with center} & \makecell{$\theta_A=\theta_B=\theta_C=\theta_D=\pi/2$} \\
\hline
\multirow{4}{*}{\centering $J>1$} & Parallel or Prism & \begin{tabular}[c]{@{}c@{}}$\theta_A+\theta_C=\pi$\\ $\theta_B=\theta_D=\pi/2$\end{tabular} \\
\cline{2-3}
&  Prism with center & \begin{tabular}[c]{@{}c@{}}$\theta_A=\pi/2$\\ $\theta_B+\theta_C+\theta_D=3\pi/2$\\ $\theta_i=\pi/2$, ($i=B,C$ or $D$)\end{tabular}\\
\hline
\end{tabular}
\label{tab:tablen4rules}
\end{table}

It is also worth emphasizing that for \( j=2 \), the values of \( \theta_A \) and \( \theta_C \) in the \( Parallel \) and \( Prism \) configurations with four particles are slightly different, whereas they are identical for the case of three particles, as presented in Appendix \ref{appendix:B}.

\begin{figure}[htbp]  
    \centering
    \includegraphics[width=0.9\textwidth]{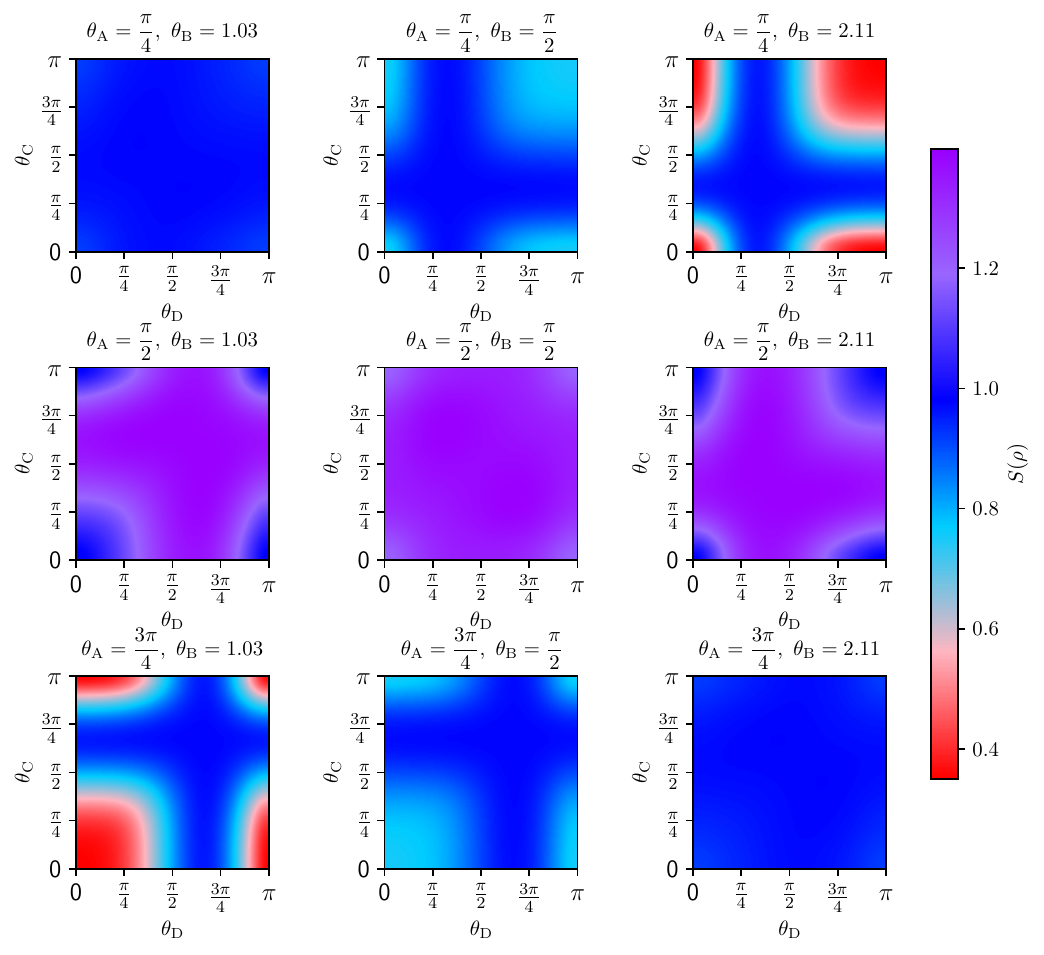}  
    \captionsetup{justification=raggedright, singlelinecheck=false}
    \caption{
    The contour plot for the entanglement entropy over $(\theta_C,\theta_D)$ plane for the \( Prism\ with\ center \) configuration with four particles with \( j=2 \) at $t=2$s.} 
    \label{fig:n4j29plot} 
\end{figure}

Next, we focus on the \( Prism\ with\ center \) configuration  and  examine the behavior of the entanglement entropy on ($\theta_C, \theta_D$) plane with fixed angles \( \theta_A \) and \( \theta_B \), as illustrated in Fig. (\ref{fig:n4j29plot}). It is evident that when \( \theta_A \) deviates from \( \pi/2 \), the entropy decreases significantly. For instance, in the middle column of Fig. (\ref{fig:n4j29plot}), where \( \theta_A \) varies from \( \pi / 4 \), to \( \pi / 2 \), and then to \( 3\pi / 4 \), the color transitions from blue to purple and then back to blue, demonstrating a marked decline in entropy as \( \theta_A \) departs from \( \pi / 2 \). Furthermore, when \( \theta_A = \theta_B = \pi / 2 \), variations in \( \theta_C \) and \( \theta_D \) have a negligible effect on the entropy. This indicates that the remaining angles play a minor role in altering the entanglement entropy under these conditions. Additionally, with \( \theta_A \) fixed, changes in \( \theta_B \) exert only a minor influence on the entropy, as evident in any row of Fig. (\ref{fig:n4j29plot}). In addition, the results for the cases \( (\theta_A, \theta_B) \) and \( (\pi - \theta_A, \pi - \theta_B) \) exhibit a strong symmetry. This symmetry arises naturally from the geometric symmetry of the \( Prism\ with\ center \) configuration.

Finally, we are concerned with the time evolution of the entanglement entropy for different configurations of the system with four particles, as shown in Fig. (\ref{fig:n4time}), while the complete period is illustrated in Appendix \ref{appendix:B}. From Fig. (\ref{fig:subn4time1}), it is evident that given the same spin $j$, the \( Prism\ with\ center \) configuration achieves the largest entanglement entropy among all the configurations. Moreover, as the spin increases, the maximum entropy attained by this configuration also increases. 

Fig. (\ref{fig:subn4time2}) reveals that even for large spin values, increasing the number of particles leads to an increase in both the maximum entanglement entropy and the entropy generation rate. Notably, the increment in maximum entropy from \( N=2 \) to \( N=3 \) is larger than that from \( N=3 \) to \( N=4 \), although the entropy generation rate continues to increase consistently across all cases.

\begin{figure}[htbp]
    \centering
    \begin{subfigure}[t]{0.48\textwidth}
        \centering
        \includegraphics[width=\textwidth]{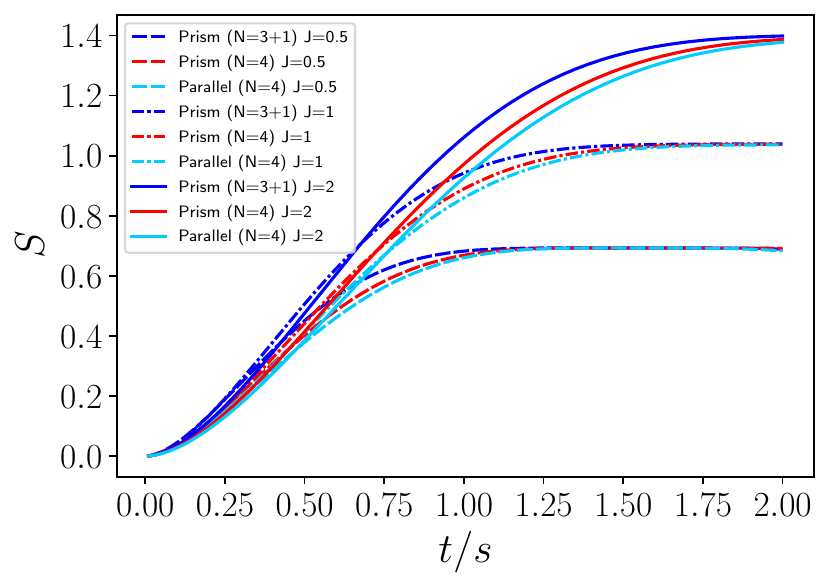}
        \caption{The evolution of different configurations}
        \label{fig:subn4time1}
    \end{subfigure}
    \hfill 
    \begin{subfigure}[t]{0.48\textwidth}
        \centering
        \includegraphics[width=\textwidth]{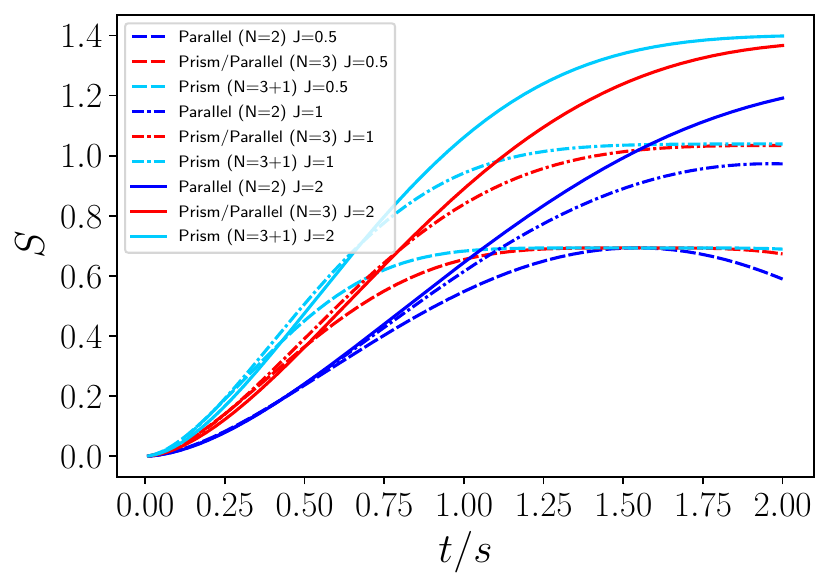}
        \caption{The evolution of different spins and particles}
        \label{fig:subn4time2}
    \end{subfigure}
    \captionsetup{justification=raggedright, singlelinecheck=false}
    \caption{
    Time evolution of the entanglement entropy  for various configurations and particle numbers.
}
    \label{fig:n4time}
\end{figure} 

\newpage

\section{Decoherence} \label{sec:decoherence}
So far, we have assumed the system to be closed, neglecting any interaction with the environment. Considering that our goal is to optimize the experimental setup for a faster rate of entanglement generation, and recognizing that entanglement is a fragile resource easily destroyed by environmental effects, in this section, we discuss the impact of increasing the particle number on decoherence.

The numerous sources of decoherence have been studied extensively \cite{2014arXiv1404.2635S,2007,Rijavec:2020qxd,Toros:2020dbf,Toros:2020krn,Fragolino:2023agd}. Although the exact nature of decoherence depends on specific environmental interactions, a general formalism can still be employed for its analysis, as introduced in the theory of decoherence presented in Ref. \cite{2007}. For a superposition state with a fixed spatial width of $\Delta x$, it was found in Ref. \cite{2007} that decoherence induced by particle scattering can be categorized into two subgroups: the short-wavelength limit and the long-wavelength limit. The total decoherence rate is given by $\gamma_{tot} = \gamma_{short} + \Gamma_{long} \Delta x ^2$, where $\gamma_{short}$ is the short-wavelength decoherence rate (e.g., from scattering by air molecules) in units of Hz, and $\Gamma_{long}$ is the long-wavelength decoherence rate (e.g., from scattering by blackbody photons) in units of $\text{Hz}/m^2$. Explicit expression of $\gamma_{tot}$ for the case of $J=1/2$ was provided in Refs. \cite{vandeKamp:2020rqh}. Furthermore, the latter reference \cite{Schut:2021svd} found that a setup with a larger number of particles exhibits greater resilience to decoherence. Decoherence for the case of $J > 1/2$ was investigated in Ref. \cite{Braccini:2023eyc}. It was found that for $J > 1/2$, we cannot define a single total decoherence rate $\gamma_{tot} = \gamma_{short} + \Gamma_{long} \Delta x ^2$; instead, the short-wavelength and long-wavelength limits must be analyzed separately. The detailed calculations are presented in Appendix E of Ref. \cite{Braccini:2023eyc}.

Following Ref.~\cite{Braccini:2023eyc}, we compare the impact of different spin states and configurations on the negativity in both the long-wavelength and short-wavelength limits. Since the decoherence rate in the long-wavelength regime scales as $T^9$, while in the short-wavelength regime it scales as $T^{3/2}$~\cite{2007}, lowering the experimental temperature is significantly more effective at suppressing long-wavelength decoherence (e.g., blackbody radiation) than short-wavelength decoherence. Therefore, in the following analysis, we primarily focus on the effects of short-wavelength decoherence.

\begin{figure}[htbp]
    \centering
    \begin{subfigure}[t]{0.4\textwidth}
        \centering
        \includegraphics[width=\textwidth]{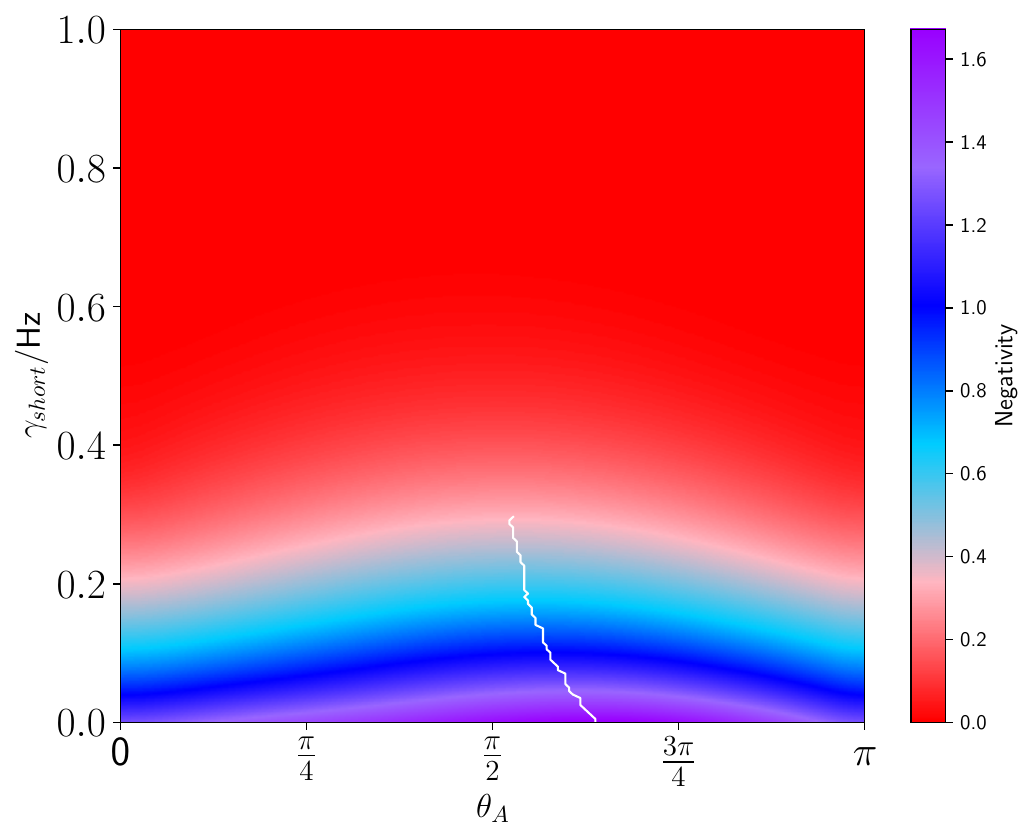}
        \caption{$J=2$ varying $\theta_A$}
    \label{fig:N3Negativity_parallel_shortwavelength}
    \end{subfigure}
    \hspace{0.05\textwidth} 
    \begin{subfigure}[t]{0.4\textwidth}
        \centering
        \includegraphics[width=\textwidth]{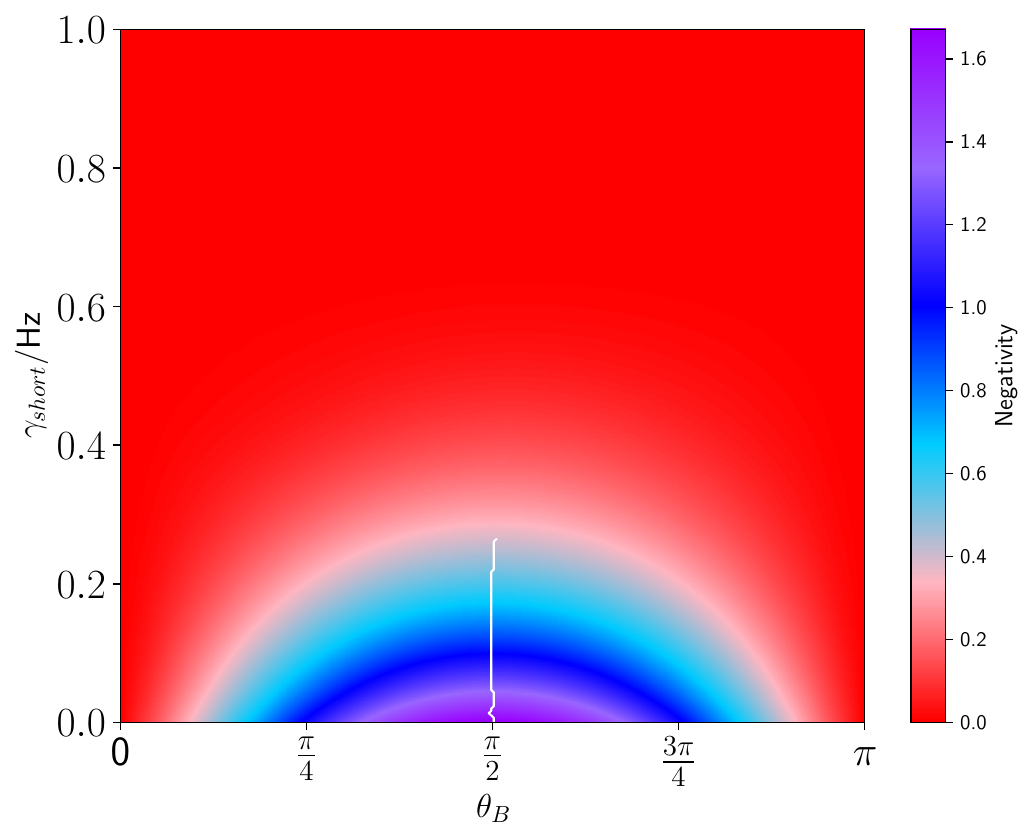}
        \caption{$J=2$ varying $\theta_B$}
        \label{fig:N4Negativity_prism_shortwavelength}
    \end{subfigure}
    \hspace{0.05\textwidth} 
    \begin{subfigure}[t]{0.4\textwidth}
        \centering
        \includegraphics[width=\textwidth]{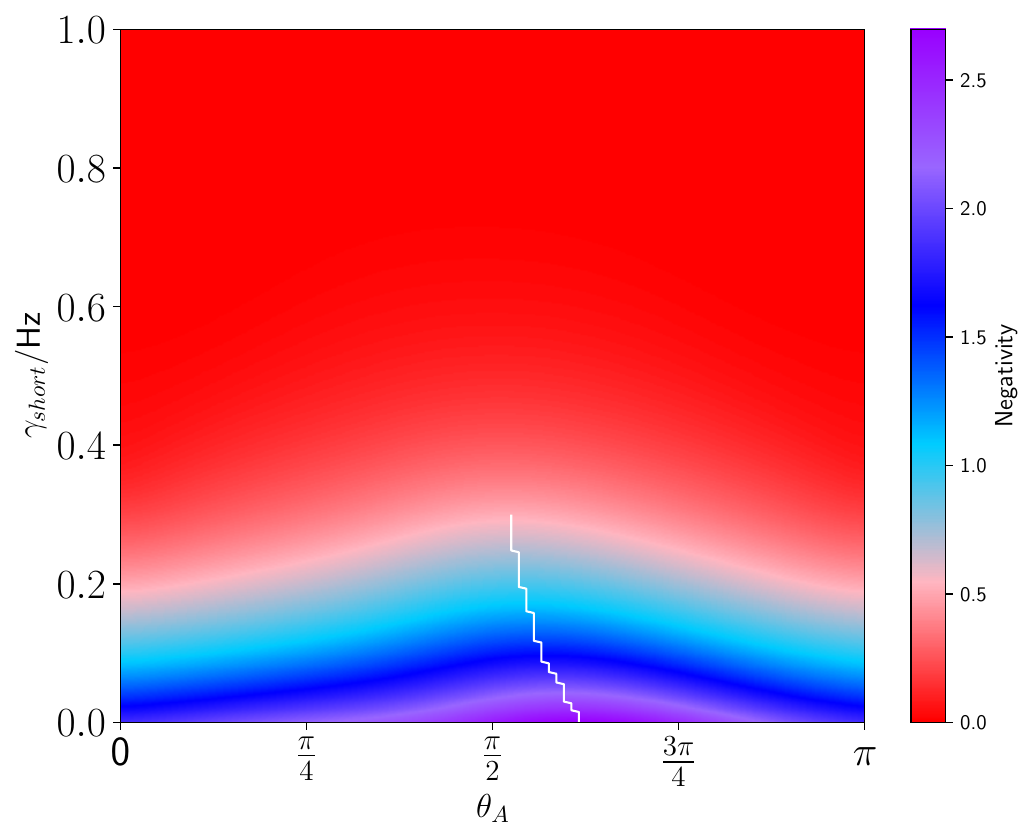}
        \caption{$J=5$ varying $\theta_A$}
        \label{fig:N4Negativity_prism_shortwavelength}
    \end{subfigure}
    \hspace{0.05\textwidth} 
    \begin{subfigure}[t]{0.4\textwidth}
        \centering
        \includegraphics[width=\textwidth]{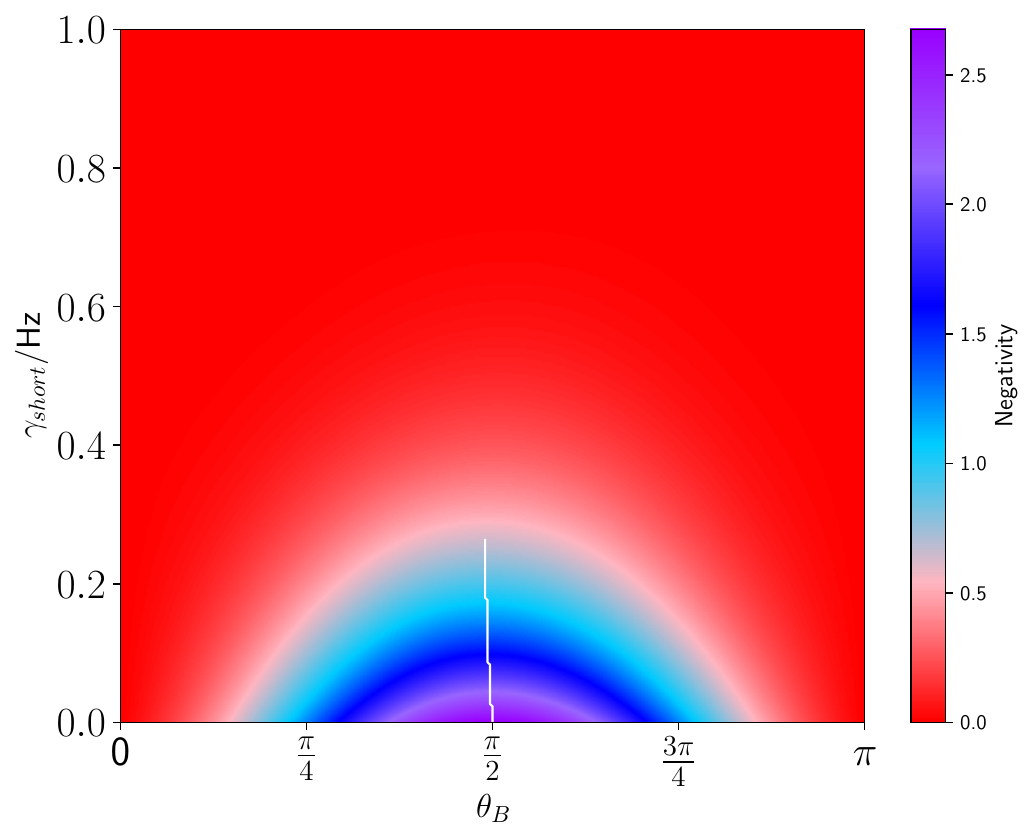}
        \caption{$J=5$ varying $\theta_B$}
        \label{fig:N4Negativity_prism_shortwavelength}
    \end{subfigure}
     \captionsetup{justification=raggedright, singlelinecheck=false}
    \caption{ Negativity as a function of $\gamma_{\text{short}}$ and $\theta_A$/$\theta_B$ for the \textit{Parallel} configuration with $N=3$ and spin values $J=2$ and $J=5$, while the other two angles are fixed at their optimal values. The white curve indicates the maximal negativity for each value of $\gamma_{\text{short}}$.}
\label{fig:n3j2j5decoherenceshort}
\end{figure} 

Assuming that the decoherence processes of individual subsystems are independent and do not influence each other, the evolution of the density matrix elements of the total system under short-wavelength decoherence can be described as follows, according to Ref.~\cite{2007}:
\begin{equation}
\rho_{m_1,m'_1,m_2,m'_2,\cdots,m_N,m'_N}^{(short)}=\rho_{m_1,m'_1,m_2,m'_2,\cdots,m_N,m'_N}e^{-(N-\delta_{m_1,m'_1}-\cdots-\delta_{m_N,m'_N})\gamma_{short}\tau},
\end{equation}
where $m_i$ and $m'_i$ are the different spin numbers of the i'$th$ subsystem, and $N$ is the total particle number of the system and $\gamma_{short}$ is the short-wavelength decoherence rate. And the density matrix elements of the total system under the long-wavelength decoherence are
\begin{equation}
\rho_{m_1,m'_1,m_2,m'_2,\cdots,m_N,m'_N}^{(long)}=\rho_{m_1,m'_1,m_2,m'_2,\cdots,m_N,m'_N}e^{-\Gamma_{long}(\Delta x)^2[(m_1-m'_1)^2+\cdots+(m_N-m'_N)^2]\tau},
\end{equation}
where $\Gamma_{long}$ is the long-wavelength decoherence rate. The explicit derivation of the density matrix under decoherence is provided in Appendix \ref{appendix:D}.

Similar to the case of the von Neumann entanglement entropy, the negativity also exhibits optimal angles at which it is maximized. In  Appendix \ref{appendix:C}, we illustrate that for the case of $N=3$ and $J=2$, in the absence of decoherence, the optimal angles for negativity coincide with those for the von Neumann entanglement entropy. 

To investigate the relationship between these optimal angles and the short-wavelength decoherence rate $\gamma_{\text{short}}$, we compute the negativity of the \textit{Parallel} configuration with $N=3$ for different spin values, $J=2$ and $J=5$, as shown in Fig. \ref{fig:n3j2j5decoherenceshort}. In this calculation, when one angle varies, the other two are fixed at the optimal values obtained in the previous section. Interestingly, we observe that the optimal angle tends to approach $\frac{\pi}{2}$ as $\gamma_{\text{short}}$ increases, which is indicated by the white curve. We confirm that a similar behavior also holds for the case of $N=4$.

\begin{figure}[htbp]
    \centering
    \begin{subfigure}[t]{0.4\textwidth}
        \centering
        \includegraphics[width=\textwidth]{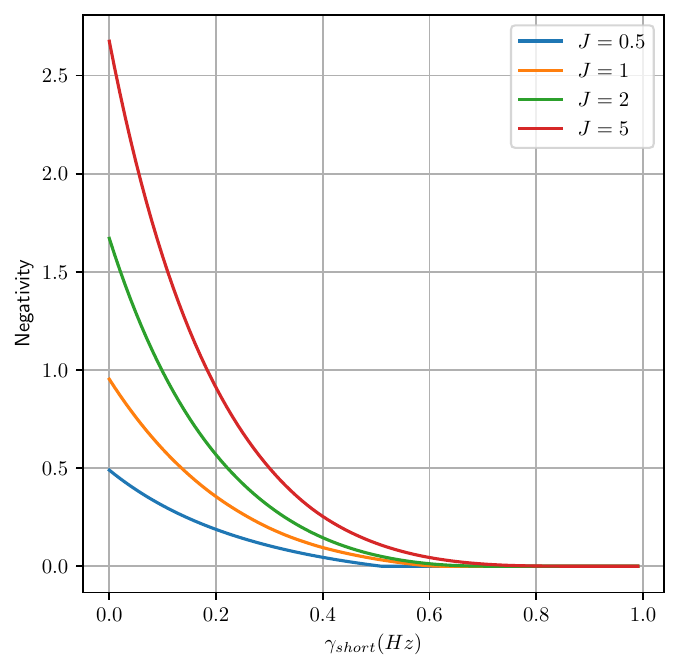}
        \caption{Short wavelength for $N=3$}
    \label{fig:N3Negativity_parallel_shortwavelength}
    \end{subfigure}
    \hspace{0.05\textwidth} 
    \begin{subfigure}[t]{0.4\textwidth}
        \centering
        \includegraphics[width=\textwidth]{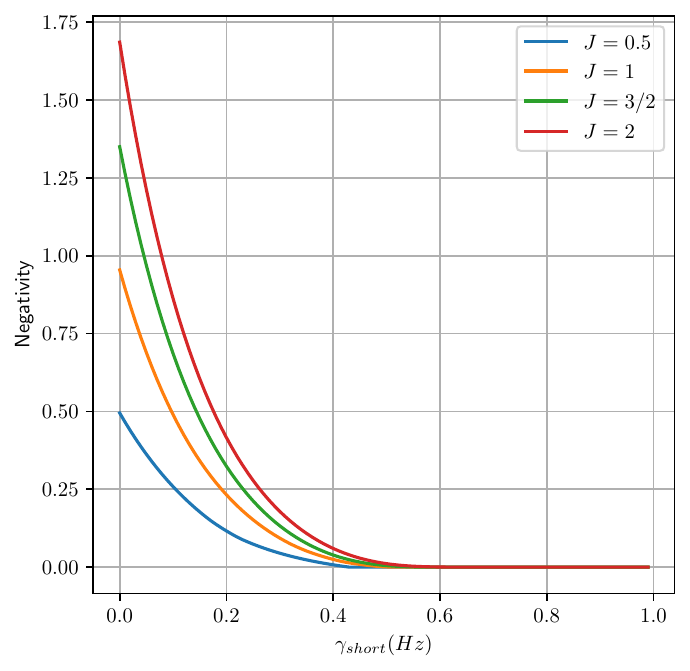}
        \caption{Short wavelength for $N=4$}
        \label{fig:N4Negativity_prism_shortwavelength}
    \end{subfigure}
    \hfill 
    \begin{subfigure}[t]{0.4\textwidth}
        \centering
        \includegraphics[width=\textwidth]{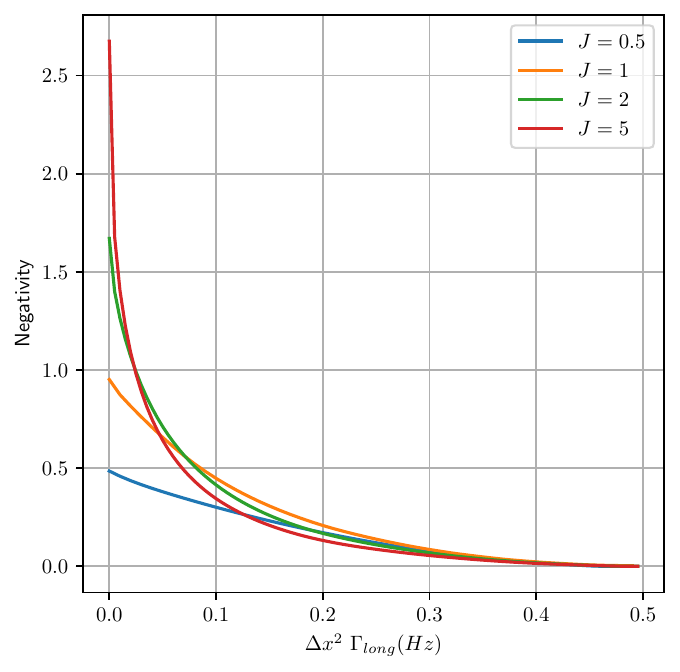}
        \caption{Long wavelength for $N=3$}
        \label{fig:N3Negativity_prism_shortwavelength}
    \end{subfigure}
    \hspace{0.05\textwidth} 
    \begin{subfigure}[t]{0.4\textwidth}
        \centering
        \includegraphics[width=\textwidth]{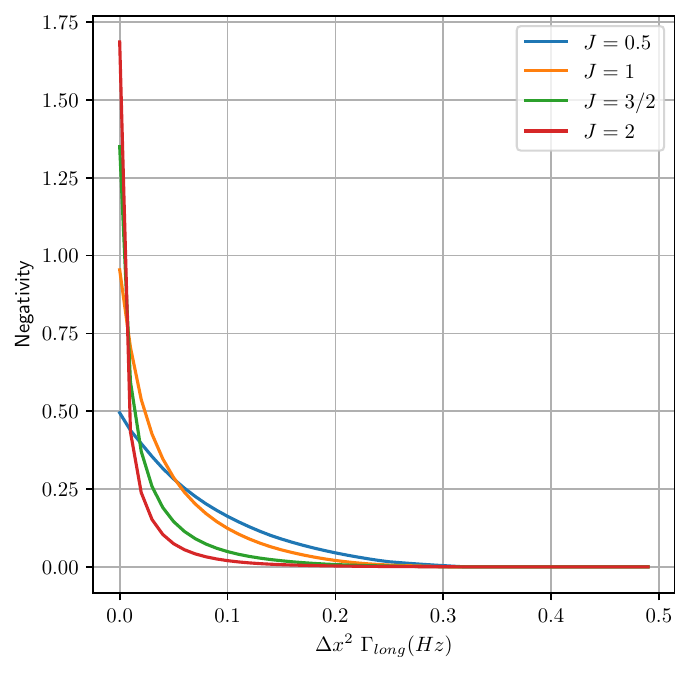}
        \caption{Long wavelength for $N=4$}
        \label{fig:N4Negativity_prism_longwavelength}
    \end{subfigure}
     \captionsetup{justification=raggedright, singlelinecheck=false}
    \caption{Negativity for different spins values as a function of the decoherence rate of \textit{Parallel} configuration for (a) short-wavelength $N=3$, (b) short-wavelength $N=4$, (c) long-wavelength $N=3$ and  (d) long-wavelength $N=4$.}
    \label{fig:n3n4shortlonggamma}
\end{figure}

From Fig. \ref{fig:n3j2j5decoherenceshort}, we also observe that if the optimal angle—obtained in the absence of decoherence—is 
fixed, the resulting negativity remains close to its maximum value, even under short-wavelength decoherence. Based on this observation, we evaluate the negativity for both short-wavelength and long-wavelength decoherence using the fixed optimal angle derived in the previous section. The corresponding results are presented in Fig. \ref{fig:n3n4shortlonggamma}.

As shown in Fig. \ref{fig:n3n4shortlonggamma}, increasing either the number of particles or the spin magnitude leads to an enhancement of negativity under short-wavelength decoherence, in agreement with the trends observed in Ref. \cite{Braccini:2023eyc}. In contrast, for long-wavelength decoherence, entanglement becomes significantly more suppressed as $\Gamma_{\text{long}}$ increases. However, since long-wavelength decoherence scales as $T^9$, its impact can be effectively mitigated by lowering the temperature, which is experimentally more feasible.

Therefore, in the presence of short-wavelength decoherence, increasing the number of particles and utilizing larger spin systems may offer a promising strategy for maintaining quantum entanglement in experimental implementations.

\newpage
\newpage
\section{Conclusion and discussion}\label{sec6}
In this paper we have investigated the gravity induced entanglement entropy in a system with multiple massive particles with large spin, which could be viewed as an improvement of the work in \cite{Li:2022yiy,Braccini:2023eyc}. Specifically, we have computed the entanglement entropy for the system with three and four particles, respectively, and figured out the optimal angles for the maximal entropy for all the allowable configurations with symmetry up to spin $j=5$ for $N=3$ and $j=2$ for $N=4$. The results have revealed that with the increase of the particle number and their spins, both the amount of the entanglement entropy and its generation rate may be greatly improved. Specifically, the entanglement entropy for \( j=2 \) reported in the original work \cite{Braccini:2023eyc} reaches its maximum value at approximately \( t \approx 2.3 \, \mathrm{s} \). In contrast, in our work, the entanglement entropy for \( j=2 \) achieves the same maximum value at \( t \approx 1.6 \, \mathrm{s} \) for the \( Prism/Parallel \) setup with \( N=3 \), and at \( t \approx 1.2 \, \mathrm{s} \) for the \( Prism\ with\ center \) setup with \( N=4 \). Notably, the time \( t \approx 1.2 \, \mathrm{s} \) required in the latter configuration is nearly half of the \( t \approx 2.3 \, \mathrm{s} \) reported in \cite{Braccini:2023eyc}. This demonstrates the increased efficiency of the \( Prism\ with\ center \) setup, particularly for higher particle numbers, in reaching the maximal entanglement entropy within a significantly shorter time frame. Alternatively, if the decoherence-preserving time remains to be fixed at $2.3$ seconds, the requirement for the mass of the matter wave could be relaxed from \( 10^{-14} \, \mathrm{kg} \) to a smaller value. This relaxation would significantly simplify the experimental implementation, making the setup more feasible under current technological constraints. In particular, the configuration of the prism with a particle at the center would provide the best efficiency for the entropy generation and this conclusion is the same as that obtained in \cite{Li:2022yiy} for particles with half spin. In contrast to the work in \cite{Li:2022yiy}, here we have found that the increase of particle spin is beneficial to the generation of the entanglement entropy.  On the other hand, in comparison with the work in \cite{Braccini:2023eyc}, we have found that the increase of the number of  particles is also beneficial to the generation of entanglement entropy. 

Furthermore, increasing the number of particles using our proposed configurations does not introduce significant experimental difficulty. For the three configurations shown in Fig. (\ref{fig:n4setup}), for instance, no major modifications to the setup are required. One would only need a magnetic field along the z-axis, with minor adjustments to the initial positions of the particles before they are split. The configurations in Figs. (\ref{fig:n3setup})(d) and (\ref{fig:n3setup})(e) would be more challenging to implement. For instance, achieving the configuration in Fig. (\ref{fig:n3setup})(d) would likely require a magnetic field with a radial gradient. However, according to our results, these two configurations do not yield superior performance and can therefore be disregarded. Given the significant benefits of adding more particles—for instance, as previously mentioned, achieving the same level of entanglement in nearly half the time—we believe the effort is well justified. However, the number of particles cannot be increased indefinitely. Using the configuration in Fig. (\ref{fig:n4setup})(b) as an example, if more than seven particles are employed, the distance between particles C and D would become smaller than the distance between A and C. This would reduce the minimum inter-particle distance, at which point the Casimir effect could become a significant factor, introducing experimental error. Therefore, a maximum of seven particles should be applied to the above setup with multiple particles.

Decoherence is a critical factor that must be carefully considered in QGEM experiments. On one hand, Ref.\cite{Schut:2021svd,Li:2022yiy}  suggests that setups with a greater number of particles are more resilient to decoherence. On the other hand, following the analysis on decoherence in \cite{Braccini:2023eyc}, while the dependency on the superposition distance in the long-wavelength limit leads to easier decoherence for larger spins, at shorter wavelengths, a larger superposition does not impact decoherence. It is relatively straightforward to eliminate long-wavelength interference caused by black-body noise in experiments, whereas mitigating short-wavelength interference presents a greater experimental challenge. Therefore, we believe that our setup, consisting of multiple particles with large spins, will exhibit greater resilience to decoherence. Our calculations regarding decoherence have already confirmed this point.

The experimental realization of the superposition states involving massive particles remains a significant challenge in contemporary physics. This difficulty arises primarily due to the intricate requirements for isolating such systems from environmental interactions and ensuring precise control over their quantum states. Furthermore, addressing and mitigating the effects of decoherence, which can rapidly degrade the coherence of quantum superpositions, demands further investigation and the development of advanced techniques. Achieving robust superposition states for massive particles will require substantial efforts in both experimental innovation and theoretical advancements. While this work focuses on entanglement and decoherence, a potential topic for future research is the preservation of robust superposition states for massive particles, particularly regarding the influence of large spin symmetry and spin degeneracy.

\section*{Acknowledgments}
We are very grateful to Pan Li for the collaboration at the early stage of this work, and to Wen-bin Pan for helpful discussions. We would also like to thank the anonymous reviewers for their insightful comments and suggestions, which have significantly improved this paper. This work is supported in part by the Innovative Projects of Science and Technology (E2545BU210) at IHEP, and the Natural Science Foundation of China (Grant Nos.~12035016,~12275275). It is also supported by Beijing Natural Science Foundation (Grant No.~1222031).

\newpage
\appendix
\section{Distance of different configurations}\label{appendix:A}
In this appendix we present the expressions for $R(x_k(m), x_l(n))$ in various configurations. Without loss of generality, we assume $1 \le k < l \le N$, where $N$ is the number of particles. For prism with center configuration, the particle at the center is regarded as the $0$-th particle. Note that $m$ and $n$ take values from $-j$ to $j$, so the $m$-th (or $n$-th) trajectory actually refers to the $(m + j)$-th (or $(n + j)$-th) trajectory.
\begin{align}
    R^{(\text{Linear})}(x_k(m), x_l(n))            & = (l-k)(\Delta s + 2j \Delta x) + (n-m)\Delta x,                                                                                                            \\
    R^{(\text{Parallel})}(x_k(m), x_l(n))          & = \sqrt{\left[(l-k)\Delta s\right]^2 + \left[(n-m)\Delta x\right]^2},                                                                                       \\
    R^{(\text{Prism})}(x_k(m), x_l(n))             & = \sqrt{\left[\Delta s\frac{\sin\left(\left(l-k\right)\frac{\pi}{N}\right)}{\sin\frac{\pi}{N}}\right]^2 + \left[(n-m)\Delta x\right]^2},                    \\
    R^{(\text{Prism with center})}(x_0(m), x_l(n)) & = \sqrt{\left(\Delta s\right)^2 + \left[(n-m)\Delta x\right]^2},                                                                                            \\
    R^{(\text{Prism with center})}(x_k(m), x_l(n)) & = \sqrt{\left[2\Delta s \sin\frac{\pi}{N}\frac{\sin\left(\left(l-k\right)\frac{\pi}{N}\right)}{\sin\frac{\pi}{n}}\right]^2 + \left[(n-m)\Delta x\right]^2}, \\
    R^{(\text{Star})}(x_k(m), x_l(n))              & = \sqrt{a^2 + b^2 - 2ab \cos\left(\left(l-k\right)\frac{2\pi}{N}\right)},                                                                                   \\
    R^{(\text{Polygon})}(x_k(m), x_l(n))           & = 2L \sin\left[\left(l-k\right)\frac{\pi}{N}+(n-m)\theta\right],
\end{align}
where $a=\frac{\Delta s}{2 \sin(\pi/N)} + (m+j) \Delta x$, $b=\frac{\Delta s}{2 \sin(\pi/N)} + (n+j) \Delta x$, $L=\sqrt{\frac{\Delta s^2 + 2 \Delta s \left(2j\Delta x\right) \cos(\pi/N) + \left(2j\Delta x\right)^2}{4\sin^2 (\pi/N)}}$, and $\theta=\arcsin(j \Delta x/L)$.

\section{The numerical results of Von Neumann entropy}\label{appendix:B}
The optimal angles for three particles (\( N=3 \)) with various \( j \) values are summarized in Table (\ref{tablen3morej}). The symbol ``\(\sim\)'' in the table indicates that the corresponding angle can take any value in the range \([0, \pi]\). From the table, it can be observed that for the \( Linear \) and \( Polygon \) configurations, the rules vary with different spin \( j \). In contrast, for the \( Star \) and \( Prism/Parallel \) configurations, the rules remain unchanged. 
Quite interestingly, the rules governing the optimal angles partially reflect the underlying geometric symmetries of the configurations. For instance, the \textit{Star}, \textit{Parallel}, and \textit{Prism} configurations in Fig.~(\ref{fig:n3setup}) are invariant under a mirror reflection through the axis containing particle~$B$. This symmetry corresponds to the exchange of particles~$A$ and~$C$, implying that the angles $\theta_A$ and $\theta_C$ are interchangeable, as confirmed by the results in Table~(\ref{tablen3morej}). In contrast, the \textit{Linear} and \textit{Polygon} configurations lack such mirror symmetry, and therefore $\theta_A$ and $\theta_C$ are not equivalent, as clearly illustrated in Table~\ref{tablen3morej} for $J > \tfrac{1}{2}$.

\begin{table}[]
\caption{The optimal angles for the maximal entropy with $N=3$ for various $J$ values}
\label{tablen3morej}
\begin{tabular}{|c|c|c|c|c|c|}
\hline
Configuration          & S($t$ = 2s) & $\theta_A$      & $\theta_B$    & $\theta_C$      & Rules                                                                                             \\ \hline
Linear (J=1/2)         & 0.638   & 1.57        & 1.57      & 1.57        & \multirow{5}{*}{$\theta_A=\theta_B=\theta_C=\frac{\pi}{2}$}                                                        \\ \cline{1-5}
Polygon (J=1/2)        & 0.638   & 1.57        & 1.57      & 1.57        &                                                                                                   \\ \cline{1-5}
Star (J=1/2)           & 0.640   & 1.57        & 1.57      & 1.57        &                                                                                                   \\ \cline{1-5}
Prism/Parallel (J=1/2) & 0.673   & 1.57        & 1.57      & 1.57        &                                                                                                   \\ \cline{1-5}
Prism/Parallel (J=1)   & 1.034   & 1.57        & 1.57      & 1.57        &                                                                                                   \\ \hline
Linear (J=1)           & 0.802   & 1.15/1.99   & 1.57      & 1.99/1.15   & \multirow{2}{*}{\begin{tabular}[c]{@{}c@{}}$\theta_B=\frac{\pi}{2}$\\ $\theta_A+\theta_C=\pi$\end{tabular}}           \\ \cline{1-5}
Polygon (J=1)          & 0.843   & 1.15/1.99   & 1.57      & 1.99/1.15   &                                                                                                   \\ \hline
Linear (J=3/2)         & 0.588   & 0.95/2.19   & 1.16/1.98 & 2.19/0.95   & \multirow{2}{*}{$\theta_A+\theta_C=\pi$}                                                                 \\ \cline{1-5}
Polygon (J=3/2)        & 0.615   & 0.95/2.19   & 1.16/1.98 & 2.19/0.95   &                                                                                                   \\ \hline
Linear(J=2)            & 0.540   & 2.31/$\sim$ & 0.83/2.31 & $\sim$/0.83 & \multirow{4}{*}{\begin{tabular}[c]{@{}c@{}}$\theta_A+\theta_B=\pi$\\ or\\ $\theta_B+\theta_C=\pi$\end{tabular}} \\ \cline{1-5}
Polygon(J=2)           & 0.566   & 2.31/$\sim$ & 0.83/2.31 & $\sim$/0.83 &                                                                                                   \\ \cline{1-5}
Linear(J=5)            & 0.538   & 2.61/$\sim$ & 0.53/2.61 & $\sim$/0.53 &                                                                                                   \\ \cline{1-5}
Polygon(J=5)           & 0.563   & 2.61/$\sim$ & 0.53/2.61 & $\sim$/0.53 &                                                                                                   \\ \hline
Star (J=1)             & 0.649   & 1.15        & 1.15      & 1.15        & \multirow{5}{*}{$\theta_A=\theta_B=\theta_C$}                                                             \\ \cline{1-5}
Star (J=3/2)           & 0.652   & 0.95        & 0.95      & 0.95        &                                                                                                   \\ \cline{1-5}
Star(J=2)              & 0.653   & 0.83        & 0.83      & 0.83        &                                                                                                   \\ \cline{1-5}
Star (J=5)             & 0.655   & 0.53        & 0.53      & 0.53        &                                                                                                   \\ \hline
Prism/Parallel (J=3/2) & 1.237   & 1.25/1.89   & 1.57      & 1.89/1.25   & \multirow{5}{*}{\begin{tabular}[c]{@{}c@{}}$\theta_B=\frac{\pi}{2}$\\ $\theta_A+\theta_C=\pi$\end{tabular}}           \\ \cline{1-5}
Prism/Parallel(J=2)    & 1.367   & 1.17/1.97   & 1.57      & 1.97/1.17   &                                                                                                   \\ \cline{1-5}
Prism/Parallel (J=3)   & 1.515   & 1.24/1.90   & 1.57      & 1.90/1.24   &                                                                                                   \\ \cline{1-5}
Prism/Parallel (J=4)   & 1.598   & 1.29/1.85   & 1.57      & 1.85/1.29   &                                                                                                   \\ \cline{1-5}
Prism/Parallel(J=5)    & 1.647   & 1.34/1.80   & 1.57      & 1.80/1.34   &                                                                                                   \\ \hline
\end{tabular}
\end{table}

\begin{figure}[htbp]
    \centering
    \begin{subfigure}[t]{0.48\textwidth}
        \centering
        \includegraphics[width=\textwidth]{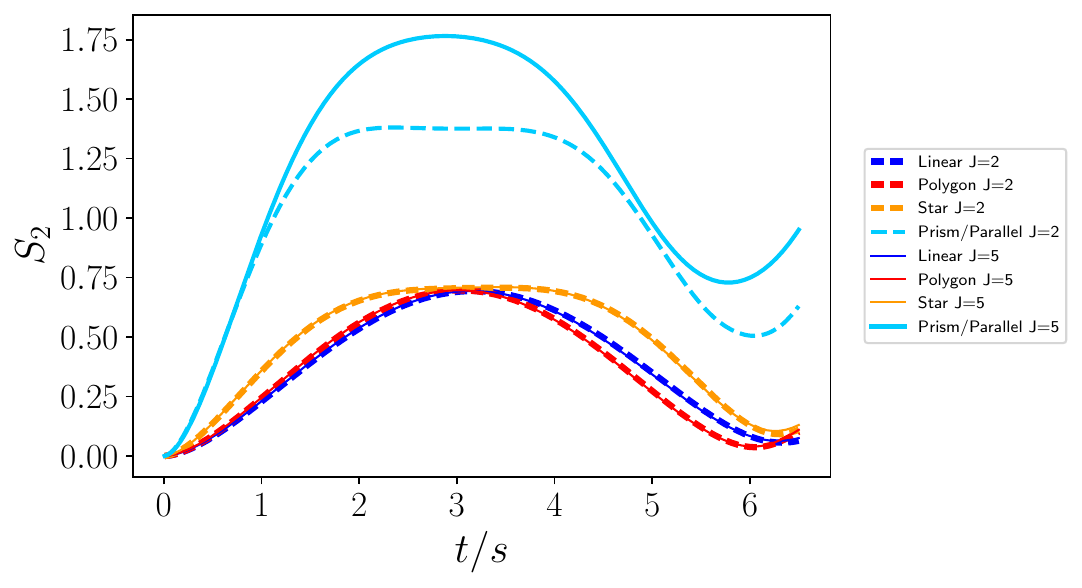}
        \caption{The evolution of different configurations}
        \label{fig:n3timeevolutionlonger1}
    \end{subfigure}
    \hfill 
    \begin{subfigure}[t]{0.48\textwidth}
        \centering
        \includegraphics[width=\textwidth]{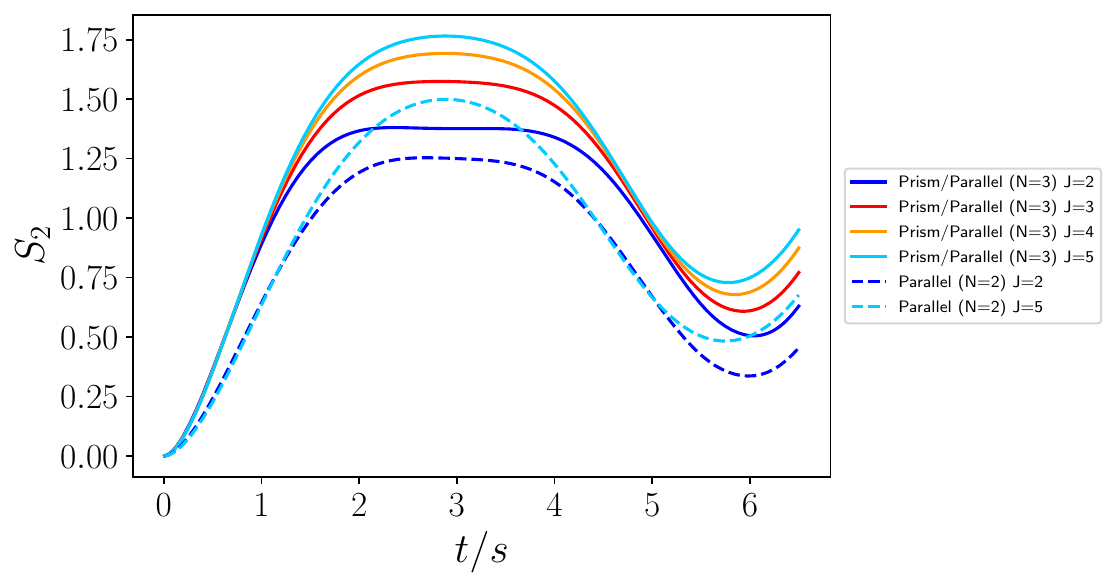}
        \caption{The evolution of different spins and particles}
        \label{fig:n3timeevolutionlonger2}
    \end{subfigure}
     \captionsetup{justification=raggedright, singlelinecheck=false}
    \caption{ Time evolution for longer time of the von Neumann entanglement entropy with three particles for various configurations and spins. 
}
    \label{fig:n3timeevolutionlonger}
\end{figure}

\begin{figure}[!ht]
    \centering
    \begin{subfigure}[t]{0.48\textwidth}
        \centering
        \includegraphics[width=\textwidth]{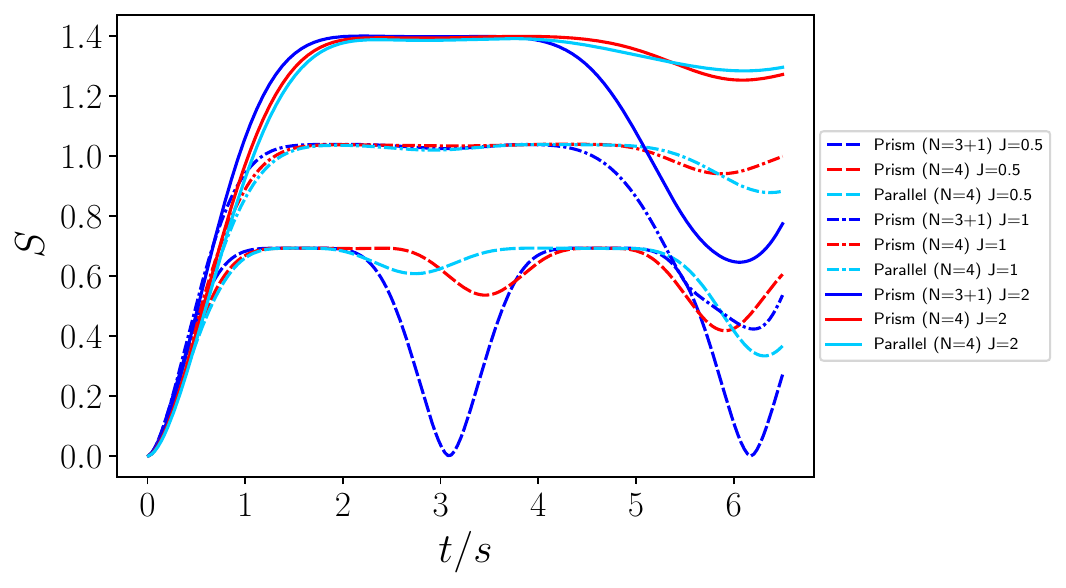}
        \caption{The evolution for different configurations}
        \label{fig:n24timeevolutionlonger1}
    \end{subfigure}
    \hfill 
    \begin{subfigure}[t]{0.48\textwidth}
        \centering
        \includegraphics[width=\textwidth]{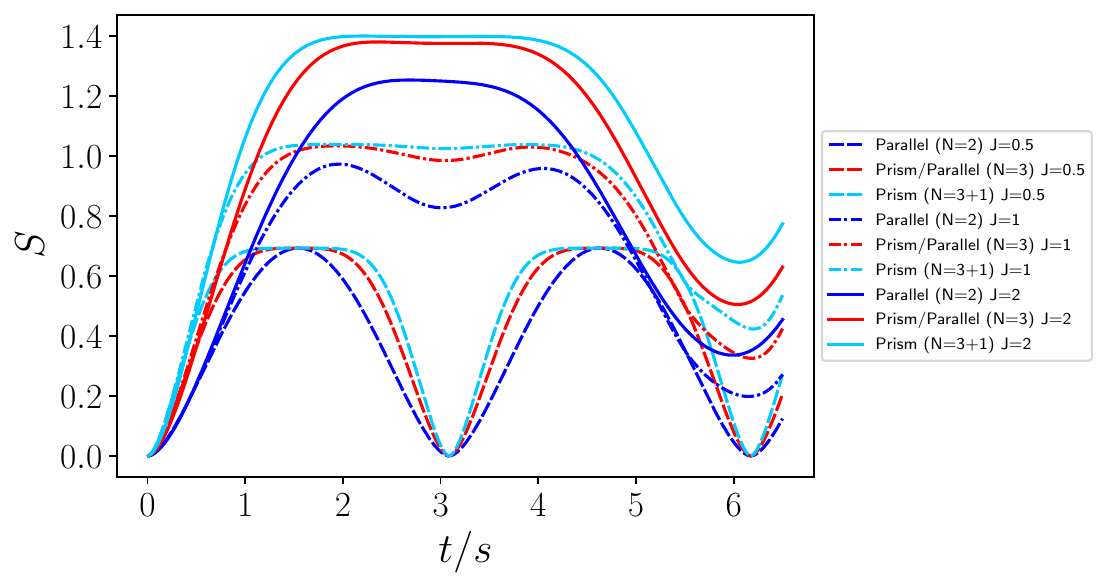}
        \caption{The evolution for different spins and particles}
        \label{fig:n24timeevolutionlonger2}
    \end{subfigure}
     \captionsetup{justification=raggedright, singlelinecheck=false}
    \caption{ Time evolution of the von Neumann entanglement entropy for longer time for various configurations and particle numbers. 
}
    \label{fig:n24timeevolutionlonger}
\end{figure}

Furthermore, the complete periods of the evolution process for two, three, and four particles are presented in Fig. (\ref{fig:n3timeevolutionlonger}) and Fig. (\ref{fig:n24timeevolutionlonger}). From these plots, it is observed that for \( j = \frac{1}{2} \), the period is approximately \( 3 \, \mathrm{s} \) regardless of the number of particles. Similarly, for \( j > \frac{1}{2} \), the period is approximately \( 6 \, \mathrm{s} \), independent of both the particle number and the value of the spin.

We show that \( \theta_B \), \( \theta_C \), and \( \theta_D \) in the \( Prism\ with\ center \) configuration exhibit cyclic symmetry. To express this symmetry concisely, we adopt the notation ``\((,)\)'' to denote the symmetry of permutation in Table \ref{Theoptimalangleswithn4}.

\begin{table}[htpb]
\caption{The optimal angles  for the maximal entropy with $N=4$}
\label{Theoptimalangleswithn4}
\resizebox{\textwidth}{!}{
\begin{tabular}{|c|c|c|ccc|c|}
\hline
Configuration    &$S(t=2$s$)$                 & $\theta_A$                    & \multicolumn{1}{c|}{$\theta_B$} & \multicolumn{1}{c|}{$\theta_C$}    & $\theta_D$ & \multicolumn{1}{c|}{Rules}                                                                                          \\ \hline
Parallel ($J$=1/2)   & 0.683      & 1.57                      & \multicolumn{1}{c|}{1.57}   & \multicolumn{1}{c|}{1.57}      & 1.57   & \multirow{6}{*}{$\theta_A=\theta_B=\theta_C=\theta_D=\dfrac{\pi}{2}$}                                                                   \\ \cline{1-6}
Prism ($J$=1/2)      & 0.692       & 1.57                      & \multicolumn{1}{c|}{1.57}   & \multicolumn{1}{c|}{1.57}      & 1.57   &                                                                                                                     \\ \cline{1-6}
Prism with center ($J$=1/2) &0.689 & 1.57                      & \multicolumn{1}{c|}{1.57}   & \multicolumn{1}{c|}{1.57}      & 1.57   &                                                                                                                     \\ \cline{1-6}
Parallel ($J$=1)     &1.036       & 1.57                      & \multicolumn{1}{c|}{1.57}   & \multicolumn{1}{c|}{1.57}      & 1.57   &                                                                                                                     \\ \cline{1-6}
Prism ($J$=1)        &1.038       & 1.57                      & \multicolumn{1}{c|}{1.57}   & \multicolumn{1}{c|}{1.57}      & 1.57   &                                                                                                                     \\ \cline{1-6}
Prism with center ($J$=1)  &1.039 & 1.57                      & \multicolumn{1}{c|}{1.57}   & \multicolumn{1}{c|}{1.57}      & 1.57   &                                                                                                                     \\ \hline
Parallel ($J$=2)     &1.377       & 1.15/1.99                 & \multicolumn{1}{c|}{1.57}   & \multicolumn{1}{c|}{1.99/1.15} & 1.57   & \multicolumn{1}{c|}{\multirow{2}{*}{\begin{tabular}[c]{@{}c@{}}$\theta_A+\theta_C=\pi$\\ $\theta_B=\theta_D=\frac{\pi}{2}$\end{tabular}}} \\ \cline{1-6}
Prism ($J$=2)         &1.387      & 1.11/2.03                 & \multicolumn{1}{c|}{1.57}   & \multicolumn{1}{c|}{2.03/1.11} & 1.57   & \multicolumn{1}{c|}{}                                                                                               \\ \hline
Prism with center ($J$=2) &1.399  & \multicolumn{1}{c|}{1.57} & \multicolumn{3}{c|}{(1.03,1.57,2.11)}                     & \multicolumn{1}{c|}{\begin{tabular}[c]{@{}c@{}}$\theta_A=\frac{\pi}{2}$\\ $\theta_B+\theta_C+\theta_D=
\frac{3\pi}{2}$\\ $\theta_i=\frac{\pi}{2},\ (i=B\ $or$\ C\ $ or$\ D)$\end{tabular}}           \\ \hline
\end{tabular}
}
\end{table}

\newpage

\section{The numerical results of negativity}\label{appendix:C}
In this Appendix, we show the supplementary data of the negativity based on the numerical analysis.
\begin{figure}[htbp]
    \centering
    \begin{subfigure}[t]{0.3\textwidth}
        \centering
        \includegraphics[width=\textwidth]{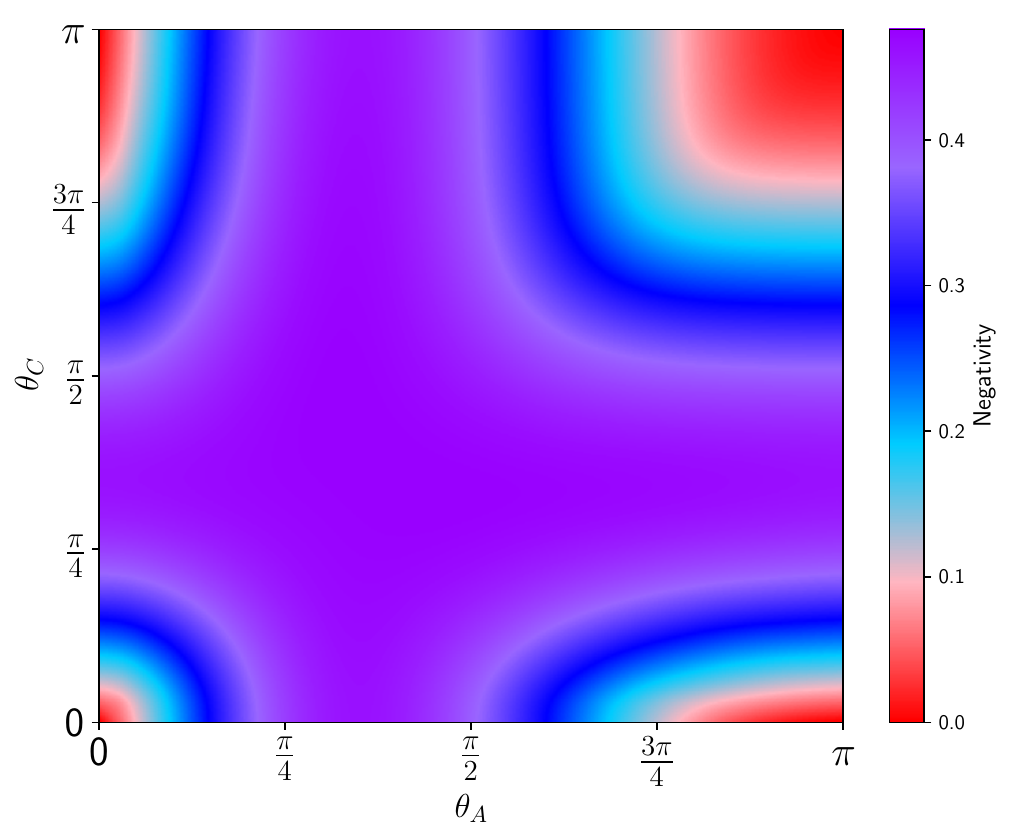}
        \caption{$\theta_B=\frac{\pi}{8}$}
    \label{fig:N3Negativity_parallel_shortwavelength}
    \end{subfigure}
    \hfill 
    \begin{subfigure}[t]{0.3\textwidth}
        \centering
        \includegraphics[width=\textwidth]{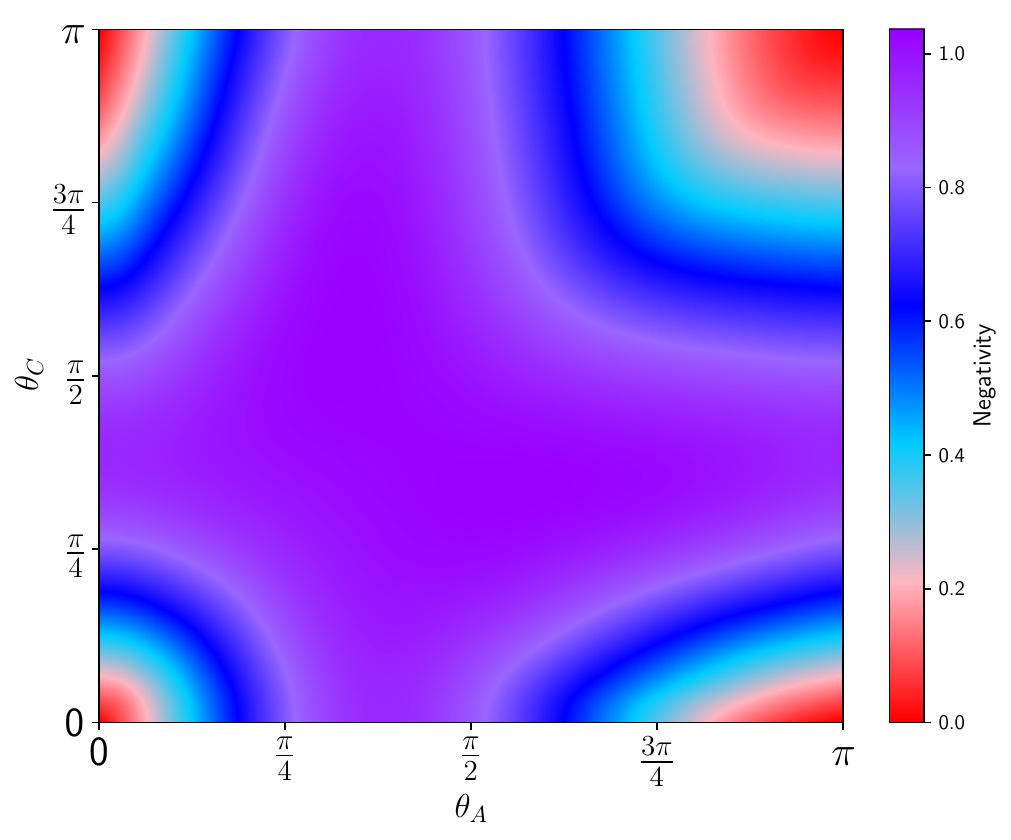}
        \caption{$\theta_B=\frac{\pi}{4}$}
        \label{fig:N4Negativity_prism_shortwavelength}
    \end{subfigure}
    \hfill 
    \begin{subfigure}[t]{0.3\textwidth}
        \centering
        \includegraphics[width=\textwidth]{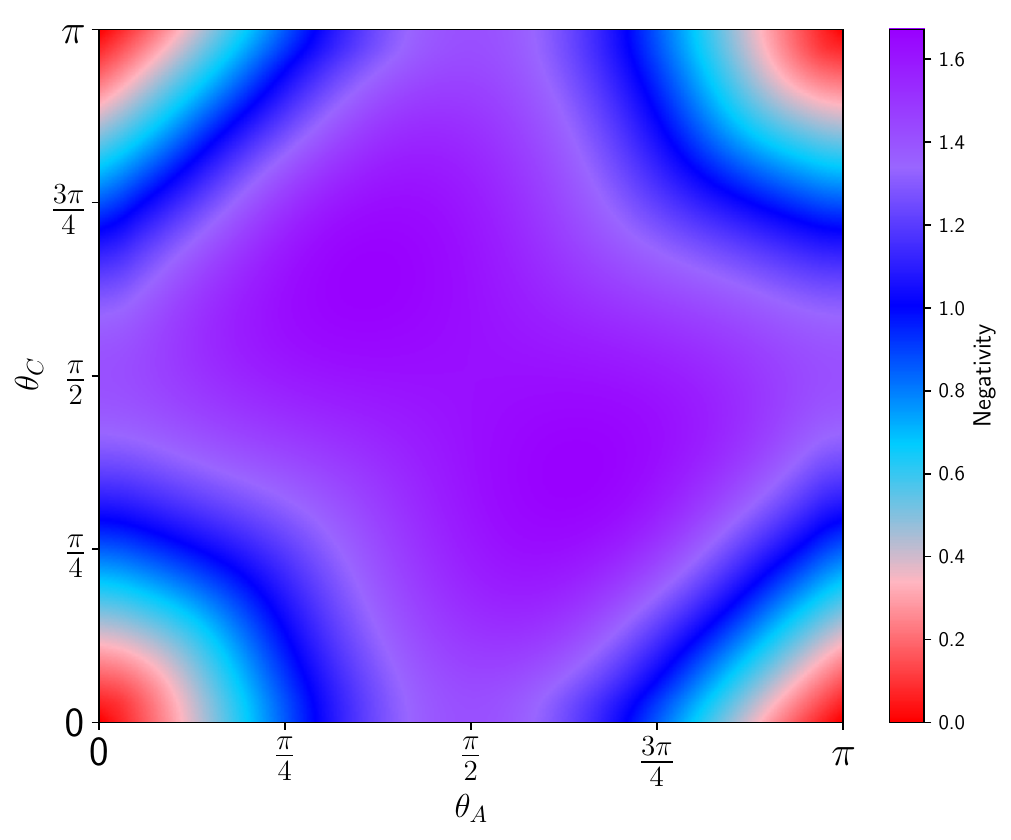}
        \caption{$\theta_B=\frac{\pi}{2}$}
    \label{fig:N3Negativity_parallel_shortwavelength}
    \end{subfigure}
     \captionsetup{justification=raggedright, singlelinecheck=false}
    \caption{The contour plot for the negativity over $(\theta_A, \theta_C)$ plane for different values of $\theta_B$ in the \textit{Parallel} configuration with three particles at t = 2s.}

\end{figure}

\begin{figure}[htbp]
    \centering
    \begin{subfigure}[t]{0.4\textwidth}
        \centering
        \includegraphics[width=\textwidth]{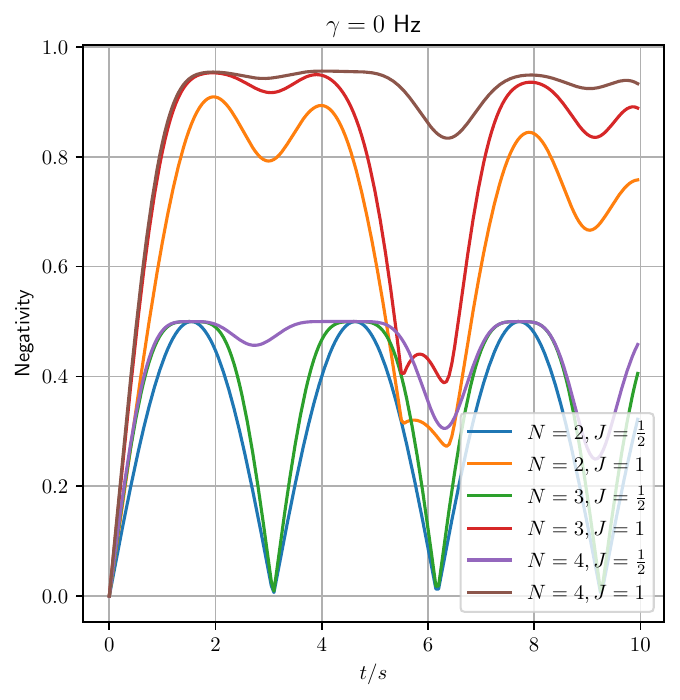}
    \label{fig:N3Negativity_parallel_shortwavelength}
    \end{subfigure}
    \hspace{0.05\textwidth} 
    \begin{subfigure}[t]{0.4\textwidth}
        \centering
        \includegraphics[width=\textwidth]{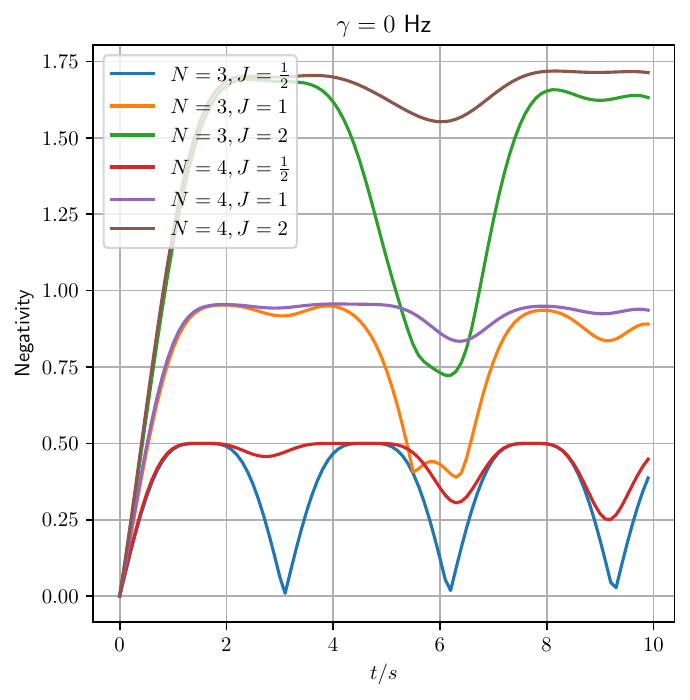}
        \label{fig:N4Negativity_prism_shortwavelength}
    \end{subfigure}
     \captionsetup{justification=raggedright, singlelinecheck=false}
    \caption{Time evolution of the negativity with two, three and four particles for various spins in the \textit{Parallel} configuration without decoherence.}

\end{figure}

\section{The density matrix under decoherence}\label{appendix:D}
In this appendix, we derive the density matrix under decoherence for both short- and long-wavelength cases. 
Following \cite{2014arXiv1404.2635S,2007}, the density matrix of a single-particle position superposition state subject to short- and long-wavelength decoherence is given by
\begin{equation}
\frac{\partial \rho^{short}(\mathbf{x},\mathbf{x'},t)}{\partial t}=-\gamma_{short} \rho(\mathbf{x},\mathbf{x'},t),\ \ \frac{\partial \rho^{long}(\mathbf{x},\mathbf{x'},t)}{\partial t}=-\Gamma_{long}(\mathbf{x}-\mathbf{x'}) ^2\rho(\mathbf{x},\mathbf{x'},t),
\end{equation}
for $\mathbf{x}\neq\mathbf{x'}$. This leads to
\begin{equation}\label{eq:sl}
\rho^{short}(\mathbf{x},\mathbf{x'},t)=\rho^{short}(\mathbf{x},\mathbf{x'},0)e^{-\gamma_{short}t},\ \ \rho^{long}(\mathbf{x},\mathbf{x'},t)=\rho^{long}(\mathbf{x},\mathbf{x'},0)e^{-\Gamma_{long}(\mathbf{x}-\mathbf{x'}) ^2t}.
\end{equation}
If we write the joint state of the system and the environment in the general form
\begin{equation}\ket{\Psi}=\ket{\mathbf{x}}\ket{E_\mathbf{x}}+\ket{\mathbf{x'}}\ket{E_\mathbf{x'}},
\end{equation}
the reduced density matrix of the system is obtained by tracing over the environment:
\begin{equation}\label{eq:generalsande}
\rho_s=Tr_e(\rho)=\sum_{i=\mathbf{x},\mathbf{x'}}\bra{E_i}\rho\ket{E_i}=\sum_{i=\mathbf{x},\mathbf{x'}}\ket{i}\bra{i}+\sum^{i\neq j}_{i,j=\mathbf{x},\mathbf{x'}}\ket{i}\bra{j}\braket{E_i|E_j}.
\end{equation}
By comparing Eq. (\ref{eq:sl}) and Eq. (\ref{eq:generalsande}), we obtain
\begin{equation}\label{eq:envdeco}
    \braket{E_\mathbf{x}|E_\mathbf{x'}}_{short}=e^{-\gamma_{short}t},\ \ \braket{E_\mathbf{x}|E_\mathbf{x'}}_{long}=e^{-\Gamma_{long}(\mathbf{x}-\mathbf{x'})^2t},\ \ \ \mathbf{x}\neq\mathbf{x'}.
\end{equation}
For the case of $N$ particles, each in a position superposition state, the joint state of the system and environment can be expressed as
\begin{equation}
\begin{aligned}
\ket{\Psi}&=\sum_{m_1,\cdots,m_N=-j,\cdots,j}\ket{\mathbf{x}(m_1)\cdots \mathbf{x}(m_N)}\ket{E_{\mathbf{x}(m_1)}\cdots E_{\mathbf{x}(m_N)}}\\
&\equiv \sum_{m_1,\cdots,m_N=-j,\cdots,j}\ket{m_1\cdots m_N}\ket{E_{m_1}\cdots E_{m_N}},
\end{aligned}
\end{equation}
where, for notational simplicity, we have defined $\ket{m_i}\equiv\ket{\mathbf{x}(m_i)}$.Tracing over the environment yields the reduced density matrix of the system:
\begin{equation}\label{eq:dmofs}
\begin{aligned}
\rho_s&=Tr_e(\ket{\Psi}\bra{\Psi})\\
&=\sum_{m_1,\cdots,m_N=-j,\cdots,j}\ket{m_1,\cdots,m_N}\bra{m_1,\cdots,m_N}\\
&\quad+\sum_{m_1,m_1',\cdots,m_N,m_N'=-j,\cdots,j}^{m_1\cdots m_N\neq m_1'\cdots m_N'}\ket{m_1\cdots m_N}\bra{m_1'\cdots m_N'}\braket{E_{m_1}\cdots E_{m_N}|E_{m_1'}\cdots E_{m_N'}}
\end{aligned}
\end{equation}

Under the assumption that the decoherence processes of different subsystems are independent and do not affect one another\footnote{A subtle point should be noted here. While in the short-wavelength regime the decoherence of each subsystem can indeed be treated as an independent process, in the long-wavelength regime one must consider the decoherence induced by a common environment. The latter case is considerably more complex, but the present treatment captures the essential characteristics of the system. Moreover, the actual decoherence under long-wavelength conditions is less severe than in the independent case\cite{PhysRevA.70.022324,10.1093/acprof:oso/9780199213900.001.0001,PhysRevE.77.011112}. Also, since it scales as $T^9$, its contribution can be safely neglected compared with that of the short-wavelength case in experiments.}, the time evolution of the density matrix can be obtained by substituting Eq. (\ref{eq:envdeco}) into Eq. (\ref{eq:dmofs}). This gives
\begin{equation}
\rho_{m_1,m'_1,m_2,m'_2,\cdots,m_N,m'_N}^{(short)}=\rho_{m_1,m'_1,m_2,m'_2,\cdots,m_N,m'_N}e^{-(N-\delta_{m_1,m'_1}-\cdots-\delta_{m_N,m'_N})\gamma_{short}\tau},
\end{equation}
and
\begin{equation}
\rho_{m_1,m'_1,m_2,m'_2,\cdots,m_N,m'_N}^{(long)}=\rho_{m_1,m'_1,m_2,m'_2,\cdots,m_N,m'_N}e^{-\Gamma_{long}(\Delta x)^2[(m_1-m'_1)^2+\cdots+(m_N-m'_N)^2]\tau}.
\end{equation}

\newpage
\bibliography{references}

\end{document}